\documentclass[a4paper,11pt]{article}
\pdfoutput=1 
\usepackage{jinstpub} 
\usepackage{lineno}
\usepackage{multirow}
\usepackage{float}
\usepackage{comment}
\usepackage{bodegraph}
\usepackage[cp1252]{inputenc}

\title{\boldmath Front-end Electronics for Timing with pico-second precision using 3D Trench Silicon Sensors}

\author[a,1]{Gian Matteo Cossu}\note{Corresponding author.}
\author[a]{and Adriano Lai}

\affiliation[a]{Istituto Nazionale Fisica Nucleare, Sezione di Cagliari, Cagliari, Italy}

\emailAdd{gianmatteo.cossu@ca.infn.it}

\abstract{The next generation of collider experiments require tracking detectors with extreme performance capabilities in terms of spatial resolution (tens of \textmu m), radiation hardness ($10^{17}~1~$MeV n$_{eq}/$cm$^2$) and timing resolution (tens of ps). 3D silicon sensors, recently developed within the TimeSPOT initiative, offer a viable solution to cope with such demanding requirements. In order to accurately characterize the timing performance of these new sensors, several read-out boards, based on discrete active components, have been designed, assembled, and tested. The same electronics is also suitable for characterization of similar pixel sensors whenever timing performance in the order and below 10 ps is a requirement. This paper describes the general characteristics needed by front-end electronics to exploit solid-state sensors with fast timing capabilities and in particular, showcases the performance of the developed electronics in the testing and characterization of fast 3D silicon sensors.}


\arxivnumber{2209.11147} 

\begin{document}

\maketitle
\flushbottom

\newpage
\section{Introduction}
\label{sec:intro}

An increasingly important requirement in experimental high energy physics concerns the need of introducing timing measurements at the level of the single pixel sensor. As an example, the Upgrade-II of the LHCb experiment at the CERN LHC has requirements of concurrent space and time resolutions of the order of 10~$\mu$m and at least 50~ps respectively at the single pixel level~\cite{FTDR}. The necessary spatial resolution can be obtained using pixels with an appropriate pitch (e.g $\sim 50\mu m$), whilst regarding the requirement on temporal resolution, different technologies for sensors have been proposed in recent years, both with and without internal gain \cite{Ferrero_2020}\cite{Iacobucci_2022}. A crucial requirement to remember is that the high constraints on spatial and temporal resolution must be maintained simultaneously alongside high radiation resistance against fluences $\phi=10^{17}~1~$MeV n$_{eq}/$cm$^2$ \cite{adriano_last}. To fulfill these extreme requests, a possibility is to use 3D pixel sensors coupled with suitable electronics \cite{3D}~\cite{Parker}. The TimeSPOT initiative \cite{timespot} (standing for TIME and SPace real-time Operating Tracker) aims at developing a full prototype detection system (based on 3D sensors) suitable for particle trackers of future particle physics experiments. The TimeSPOT project also includes the development of dedicated electronics through the production of an ASIC (Application Specific Integrated Circuit) with a 28 nm process, which includes an analog front-end and a TDC for each pixel. Pixels are arranged in a $32 \times 32$ matrix, giving a total of 1024 channels \cite{Piccolo_2022}. The final goal is to realize a final demonstrator, consisting of a complete 4D tracking system that satisfies the desired performances. The obtainable performances can be strongly limited by the characteristics of the electronics, in particular if strong constraints are present. For example, due to power dissipation capabilities, the LHCb detector has a maximum allowed power of about $1.5~\text{W}/\text{cm}^2$. 

While the integrated electronics is developed, characterization of the temporal performance with these 3D sensors has been performed using the discrete component electronics described in this paper. These electronics were designed in order to obtain the best possible timing performance without worrying about power consumption. The aim was to understand what could be the maximum performance obtainable from the system if operated in the best possible conditions. This has been done, starting from the study of the characteristics of the input signal that the electronics receive, i.e. the current generated in the pixel by a Minimum Ionising Particle (MIP).

We therefore start with section \ref{sec:2} where some general concepts are introduced. In section \ref{sec:3}, the 3D silicon sensors developed within the TimeSPOT collaboration are presented as well as the characteristics of the signals that these sensors produce. Section \ref{sec:4} is dedicated to a description of the circuit topology chosen for the analog front-end, the discrete active component characteristics needed and the results of Spice simulations about the expected performance of the system. In section \ref{sec:5}, the single channel and multi-channel boards that have been built are shown, whilst in section \ref{sec:6}, some examples of applications and results obtained with our boards are presented.

\section{Time resolution of the System}\label{sec:2}

In this paper, any system designed for timing consists of a solid-state sensor (or pixel) and its related electronics. Neglecting Time-to-Digital-Conversion (TDC) resolution, and assuming a suitable discrimination method to correct for time-walk variations, there are two dominant contributions to the timing resolution of a sensor+electronics system,

\begin{equation}
    \sigma_t^2 \cong \sigma_{sens}^2+\sigma_{ej}^2.
\end{equation}

$\sigma_{ej}$ is electronic jitter, and depends both on the output noise of the amplification stage  $\sigma_v$ and on the slope of the signal at the chosen discrimination threshold $V{'_{thr}}$,

\begin{equation}\label{jitter}
    \sigma_{ej} \equiv \frac{\sigma_v}{V'_{thr}}.
\end{equation}

The term $\sigma_{sens}$ represents contributions to the timing resolution due to the pixel. Our aim is to design a front-end electronics stage that can minimize both contributions ($\sigma_{ej}$ and $\sigma_{sens}$) to the timing resolution $\sigma_{t}$. 

The effect of the system on the input signal given by the sensor current $I_D(t)$ (assuming such system as linear) can be represented by its transfer function. This is normally expressed in the Laplace domain as a function $R_m(s)$ such that,

\begin{equation}\label{trans_func}
    V_{out}(s)= I_D(s) R_m(s),
\end{equation}

where $V_{out}(s)$ and $I_D(s)$ are the Laplace transforms of the time domain signals $V_{out}(t)$ and $I_D(t)$ respectively.
The timing performances of the system are measured through the output voltages $V_{out}(t)$ and depend on both the characteristics of the input currents supplied by the sensor and on the shaping action of the electronics enclosed in the transfer function $R_m(s)$. Transfer functions commonly used for shaping are unipolar signal shaping transfer functions \cite{blum2008particle}, defined by,

\begin{equation}\label{unipolar}
    R_m(s)=  \frac{R_{m_0}}{(1+s \tau)^{n}}.
\end{equation}

The transfer function $R_m(s)$ is a Trans-impedance as it converts an input current to a voltage output, $\tau$ is its time constant, while $R_{m_0}$ is the DC-Trans-impedance of the system measured in Ohms. 

Moving to the Fourier domain with the substitution $s=j 2\pi f$  we find the frequency $f_{-3dB}$ at which the modulus of Eq. \ref{unipolar} drops by a factor $\sqrt{2}/2$. The frequency $f_{-3dB}$, will be the bandwidth of the trans-impedance $R_m(s)$ and is given by,

\begin{equation}\label{unipolar_bw}
    f_{-3dB, n}=  \frac{1}{2 \pi \tau}\big(2^{\frac{1}{n}}-1\big)^{\frac{1}{2}}.
\end{equation}

Characteristics of the output signal such as the maximum amplitude $V_{peak}$, the peaking time $T_{peak}$ and SNR (signal to noise ratio) will depend on the shape and amplitude of the input current, the $R_{m_0}$ value and the time constant $\tau$ of the system. This will be better clarified later when the theoretical estimates of the performance of the chosen circuit for the electronics developed in section \ref{teoretical} are shown.

\section{TimeSPOT 3D silicon sensors}
\label{sec:3}
\begin{figure}[h!]
    \centering
    \includegraphics[width=0.95\textwidth]{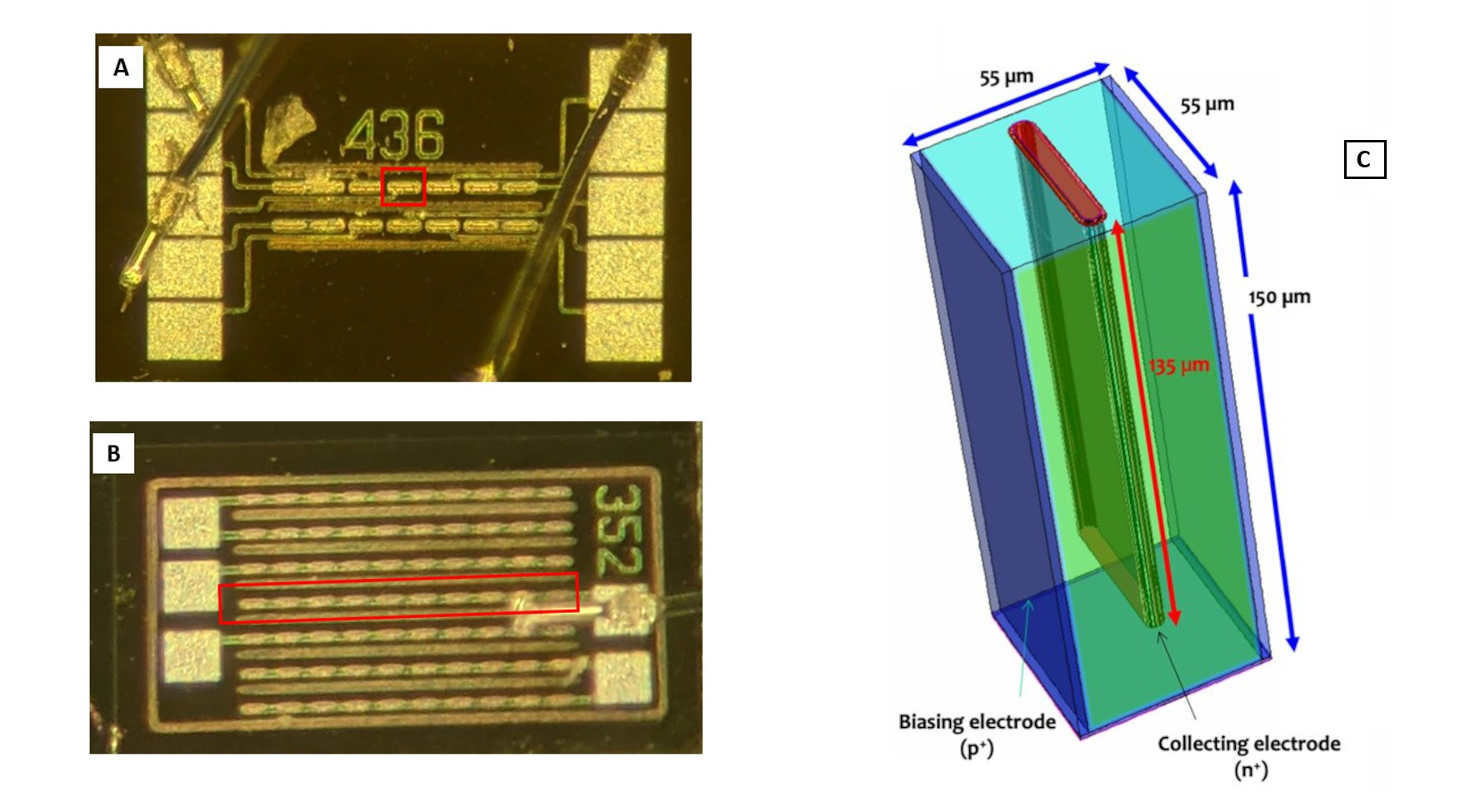}
    \caption{\footnotesize{3D-trench sensors fabricated at Fondazione Bruno Kessler (FBK, Trento, Italy) within the TimeSPOT project: A) single pixel sensor, B) strip sensors with multiple single pixel connected together, C) geometry and size of the single pixel.}}
    \label{fig:test_struct}
\end{figure}

Different solutions for solid-state planar sensor geometries have been proposed over the years, with or without gain mechanisms \cite{Ferrero_2020}\cite{Iacobucci_2022} and showing very good timing performance. An interesting alternative to such sensors is to consider a pixel with vertical (3D) geometry~\cite{3D}~\cite{Parker}. This solution completely decouples the thickness of the pixel from its inter-electrode distance. The former establishes the amount of charge produced by a MIP by linear energy deposit, whilst the latter determines the duration of the induced current signal. This feature can give the decisive advantage of larger freedom in the optimisation of the pixel geometry towards maximum timing performance. Such advantage was exploited in the TimeSPOT project~\cite{timespot}, which developed a specific and optimised 3D silicon sensor, characterized by a trench geometry \cite{LAI2020164491} \cite{JINST-TimeSpot} \cite{3D-accurate} \cite{3Dstepper}. The optimised geometry of the single pixel is shown in Fig.~\ref{fig:test_struct} c). The pixel features a pitch of $55 \mu{\rm m}$ \cite{3D-accurate}. Two ohmic trenches (represented in blue in Fig.~\ref{fig:test_struct} c) of dimensions 2.5x55x150$\,\mu{\rm m}^3$ are located at the two opposite sides of the pixel. A third trench of dimensions 5x40x130$\,\mu{\rm m}^3$ is placed at the pixel centre, parallel to the two ohmic electrodes and serves as readout electrode. 3D-trench sensors have been produced with several layouts: single pixels, double pixels, and strips with multiple single pixels connected together. The amount of charge carriers generated by an ionizing particle depends on the thickness of the pixel and on the Linear Energy Transfer (LET) of the particle; for one MIP on a pixel $150~\mu m$ thick, the released charge corresponds to a Most Probable Value (MPV)
\begin{equation} \label{cap}
Q_{MPV} \sim 2 ~\text{fC}.
\end{equation}
A key feature of a solid state sensor is its capacitance, which imposes a limit on system performance as it affects sensitivity, noise and jitter. Measurements\cite{forcolin} on the devices of the first batch produced in 2019 showed a value of the single pixel capacitance $C_D$ of
\begin{equation} \label{cap}
C_D \sim 70 \div 75 ~\text{fF}.
\end{equation}
Strip detectors have a bigger capacitance according to the number of pixel connected together. 

\begin{figure}[h!]
    \centering
    \includegraphics[scale=0.28]{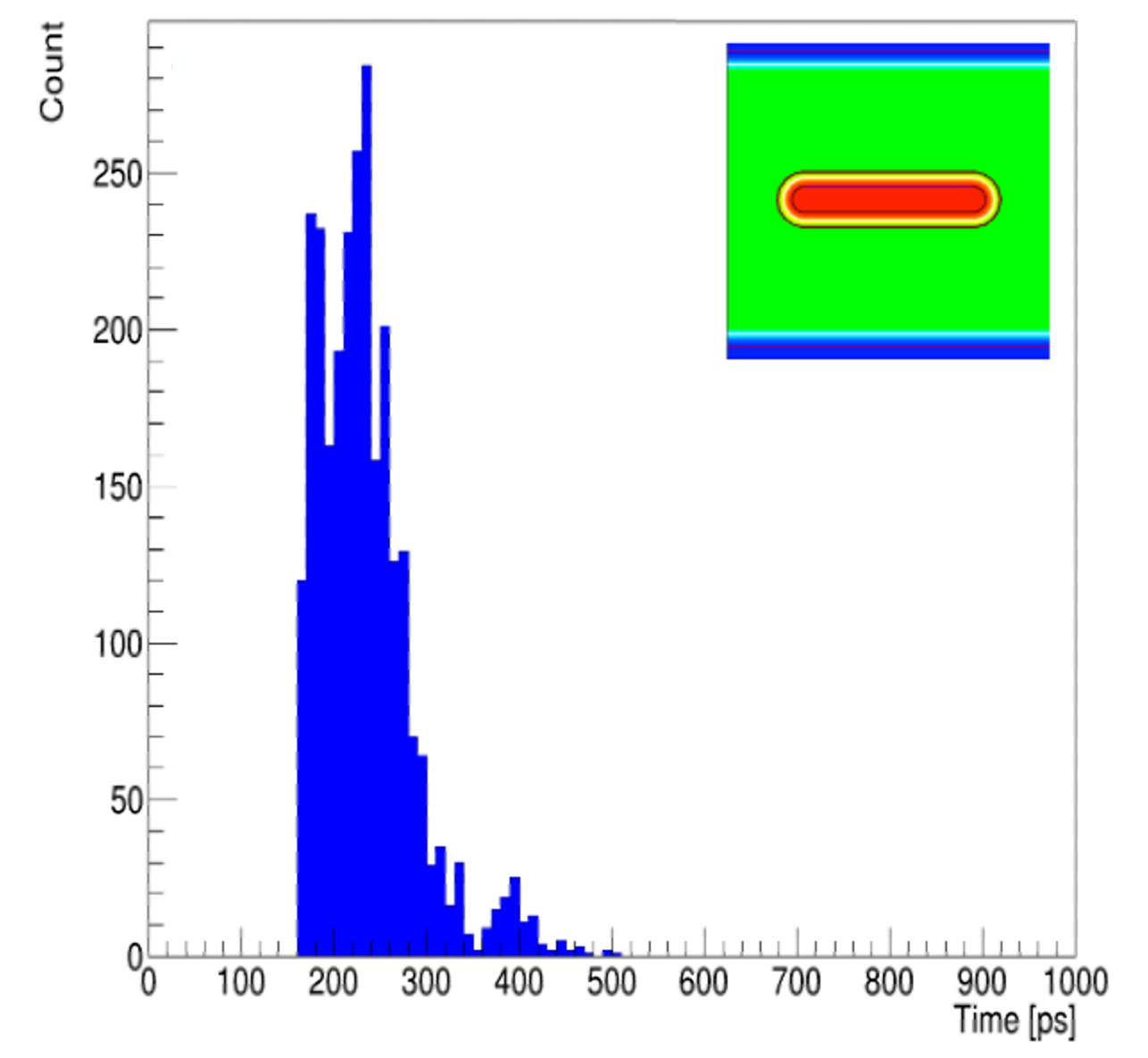}
    \caption{\footnotesize{CCT distribution of the currents of the TimeSPOT 3D trench detector obtained with the custom package TCode \cite{3D-accurate}~\cite{Loi_2021}. The average charge collection time is $\overline{t_c}=234~ \text{ps}$ while the standard deviation is $\sigma_{t_c}= 53~ \text{ps}$. }}
    \label{fig:img2}
\end{figure}

Current signals induced by ionising particles in the electrodes of this sensor type have been studied and simulated extensively using a dedicated custom software package: TCoDe \cite{3D-accurate}~\cite{Loi_2021}. In 3D geometries, the duration of current signals depends on the position of the track with respect to the electrodes.
To define the duration of the signals of the measured signals, a possible criterion is to consider the time necessary to induce a significant fraction of the total induced charge (e.g. 99.9\%). Calculating time in this way will give the collection time $t_c$ of the considered current.
The set of all collection times relating to the entire active area of the sensor can be included in a characteristic distribution, called \textit{Charge Collection Time distribution} (CCT) and shown in Fig.~\ref{fig:img2}. The CCT distribution indicates an average charge collection time $t_c \sim 200~ps$ and a standard deviation $\sigma_{t_c}\sim 50~ps$. Great effort has been made to optimize the geometry of these sensor types\cite{Loi_2021}, with the goal of reducing the variation of signal duration and shape, and minimizing the standard deviation $\sigma_{t_c}$  which is directly related to the intrinsic resolution of the sensor $\sigma_{sens}$.

\begin{figure}[h!]
    \centering
    \includegraphics[scale=0.26]{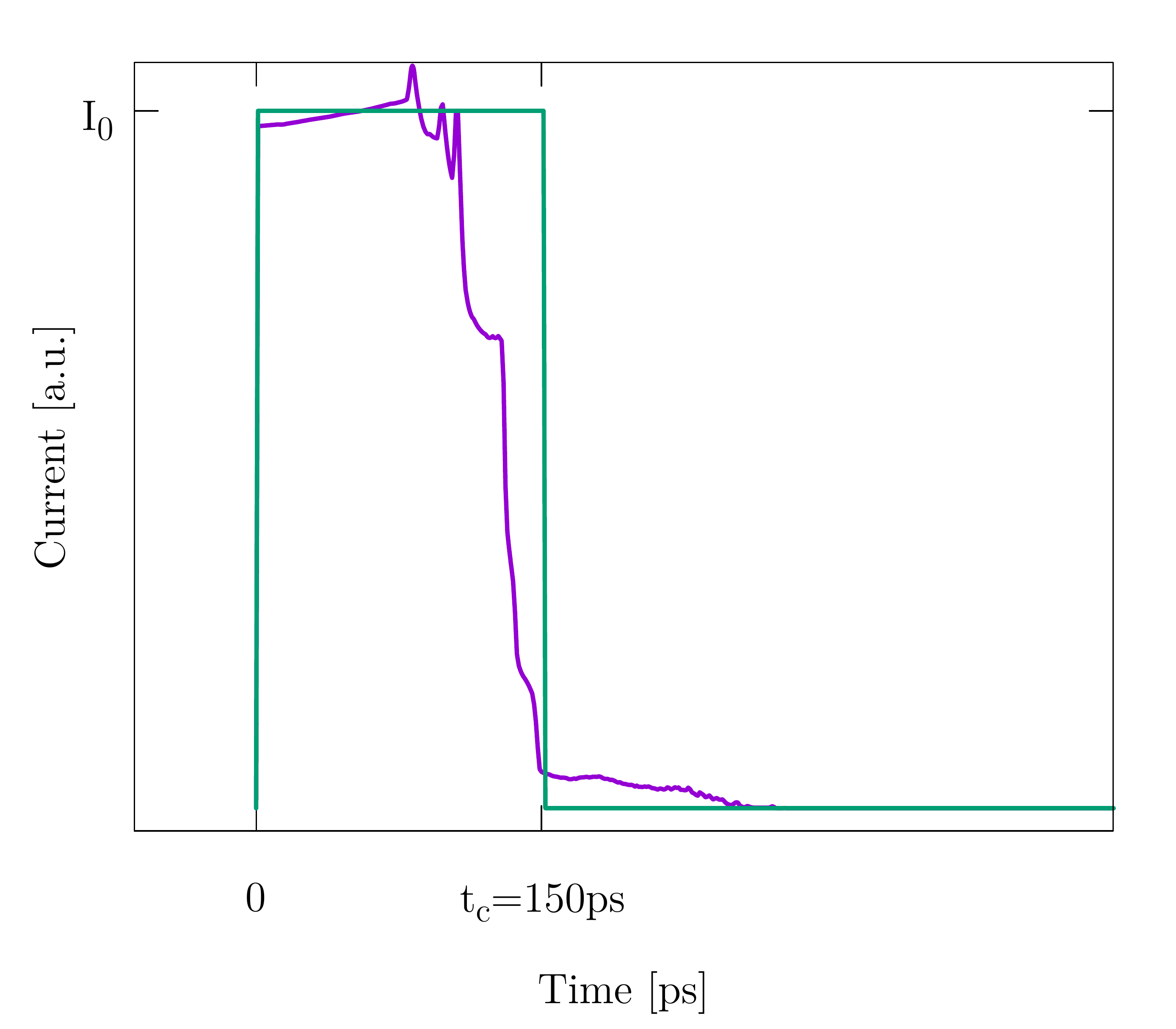}
    \includegraphics[scale=0.26]{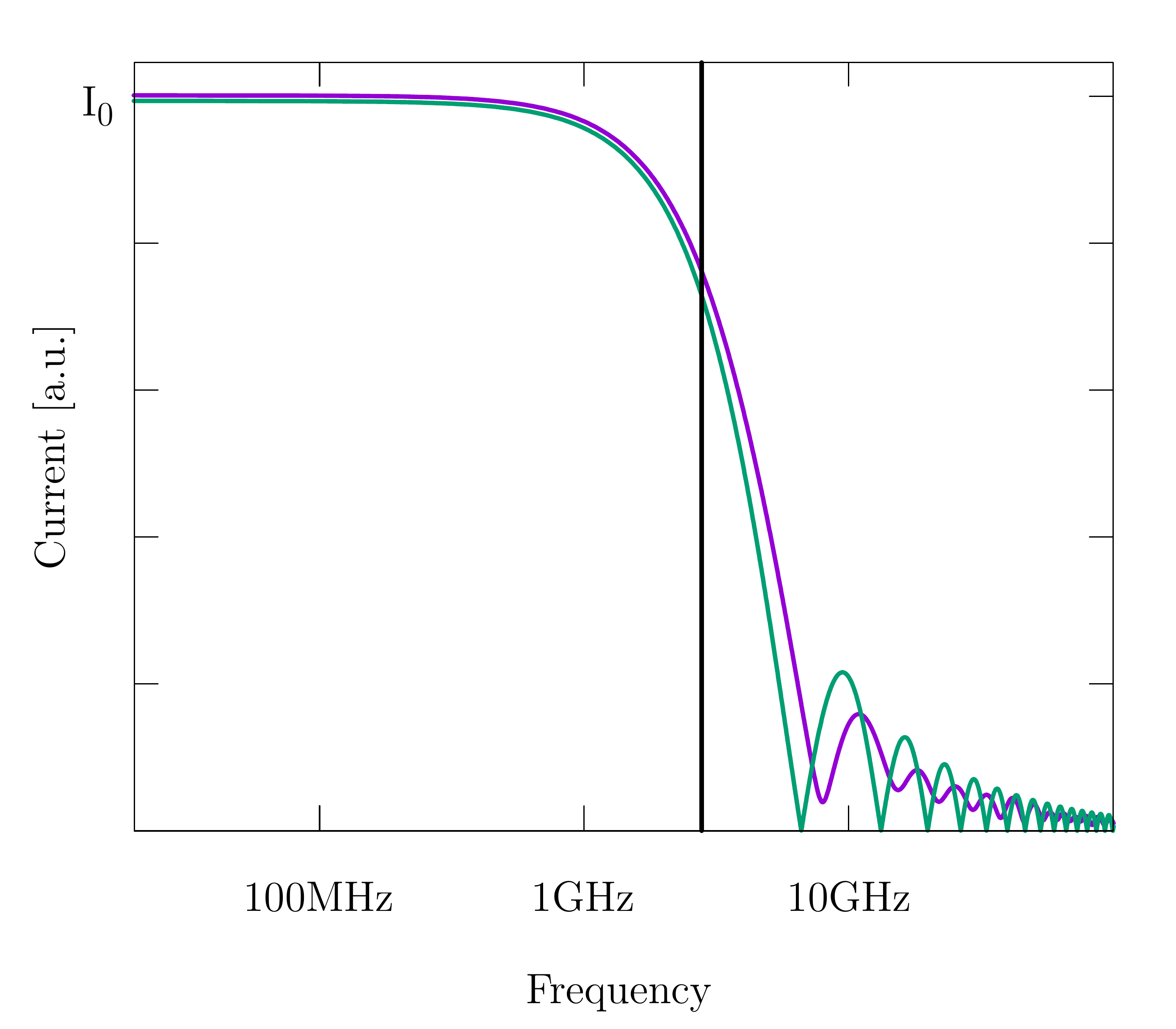}
    \caption{\footnotesize{ Figure on the left: An example of a simulated current of the single pixel obtained with TCode (purple line) and the modeled current as a rectangular pulse with duration $t_{c}$ equal to the charge collection time (green line). Figure on the right: same currents of the plot on the left but in the frequency domain: it can be seen how most of the signals is concentrated a low frequencies. The black line represent the signal bandwidth as defined in Eq. \ref{sig_band}. }}
    \label{fig:img3}
\end{figure}

An example of a typical current signal is shown in Fig.~\ref{fig:img3} (left). As a reasonable approximation, it can be assumed as a simple rectangular pulse with duration equal to the charge collection time $t_c$. 
Considering the frequency domain, the spectrum of this rectangular pulse of Fig.~\ref{fig:img3} (left), is given by the known \emph{Sinc} function: Fig.~\ref{fig:img3} (right).

The signal bandwidth is typically considered to be $f_{BW}=\frac{1}{t_c}$, and for $t_c=~200~ps$, $f_{BW}=5~GHz$, however the spectrum of the signal (Fig.~\ref{fig:img3} (right)), shows that the frequency components exhibit  gradually decreasing amplitudes as frequency increases. It is therefore reasonable to determine the signal bandwidth as the frequency at which the current spectrum amplitude drops by a factor of $\frac{\sqrt{2}}{2}$. This can be done with a second order approximation of the Sinc function, finding,

\begin{equation}\label{sig_band}
f_{-3dB} \sim \Big( \frac{6(2-\sqrt{2})}{2}\Big)^{\frac{1}{2}} \frac{1}{\pi t_c} \approx \frac{0.42}{t_c}.
\end{equation}

The average bandwidth of the signal is therefore equal to
\begin{equation}\label{sigBW}
f_{-3dB} \sim 2.1 ~\text{GHz}.
\end{equation}

The signal bandwidth found in Eq. \ref{sigBW} places strong constraints on the characteristics of the electronics in order to appreciate the speed of such sensors. As an example, a model for the electronics is considered, given by a first-order transfer function with time constant $\tau = R_{in} C_{in}$. $R_{in}$ represents the input resistance of the electronics and $C_{in}$ represents the total input capacitance given by the sum of the capacitance $C_D$ of the sensor and the one given by the electronics. The bandwidth of a first order transfer function is simply given by $f_{BW}=\frac{1}{2\pi \tau}$ (Eq. \ref{unipolar_bw}), and we want the system to have a bandwidth equal to Eq. \ref{sigBW}. If we now consider a low input resistance of $R_{in}=50~\Omega$, (this is a typical value with discrete components, especially if impedance matching is required and high frequency signals are involved ), the total input capacitance $C_{in}$ would have to be on the order of $C_{in} \approx 1~pF$ . This gives a first hint about the characteristics the input stage should have. Starting from this initial estimate, in order to obtain the best timing resolution of the system, it is now necessary to choose an appropriate circuit topology that can guarantee the best performance.

\section{Circuit description}
\label{sec:4}
This section describes the type of transistors chosen to realize the front-end amplifier and the circuit topology used. First, some of the basic characteristics of the chosen configuration will be examined, looking at the small signal model of the circuit, and finally the results of Spice-based simulations of the full circuit will be outlined.  

\subsection{Use of RF bipolar transistors in fast sensor read-out}

In the previous section we have seen the typical characteristics of signals from a 3D-trench TimeSPOT sensor. They present an average duration of 200~ps, which corresponds to a wide bandwidth in the frequency domain. In order to fully exploit the speed of this signal, a sufficiently fast active component is required. The speed of a transistor is normally characterized by its transition frequency $f_T$, i.e. the frequency at which the current gain is unity. In recent years, discrete component transistors capable of providing the best performance in terms of low noise and high transition frequency $f_T$ have been Silicon-Germanium hetero-junction bipolar transistors (HBT Si-Ge) \cite{infineon}. These devices are able to provide transition frequencies on the order of 100~GHz as well as very small values of internal capacitance, which is necessary to sustain the wide bandwidth of our sensor signals. There are several examples of the use of such devices in designing front-ends for timing applications, with planar sensors without internal gain~\cite{Benoit2016100}, and LGADs (Low Gain Avalanche Detectors)~\cite{Berretti_2017}~\cite{Minafra_2020}. In the case of LGAD readout, an excellent example is provided by the board developed at the University of California Santa Cruz~\cite{UCSC} which made it possible to obtain a timing resolution of about 30~ps~\cite{CARTIGLIA201783}. A modified version of this board, designed in the INFN laboratories at Genoa, was used for the first characterization tests on 3D TimeSPOT sensors, measuring a time resolution of about 20~ps~\cite{JINST-TimeSpot}. In the same tests it was also clearly demonstrated~\cite{JINST-TimeSpot} that the results on timing measurements were limited by the performance of the front-end electronics.

In the following sections, we demonstrate that it is possible to go further and obtain even better timing resolution from 3D TimeSPOT by using HBT Si-Ge transistors and further optimization of the read-out electronics.

\subsection{Feedback Trans-Impedance Amplifier circuit}
From the circuit point of view, the sensor is usually modeled as an ideal current generator in parallel with the capacitance $C_D$ of the detector. This generator is connected to a TIA (Trans-Impedance Amplifier) that converts the input current into an output voltage signal. The most common TIA configuration is the "voltage-current" (or "shunt-shunt") feedback topology, where a negative feedback network senses the voltage at the output and returns a proportional current to the input. This type of feedback is chosen because it lowers both the input and output resistance. A low input resistance leads the front-end to behave like a good current meter, while a low output resistance allows a better drive capability.
\begin{figure}[h]
	\centering
	\includegraphics[scale=0.39]{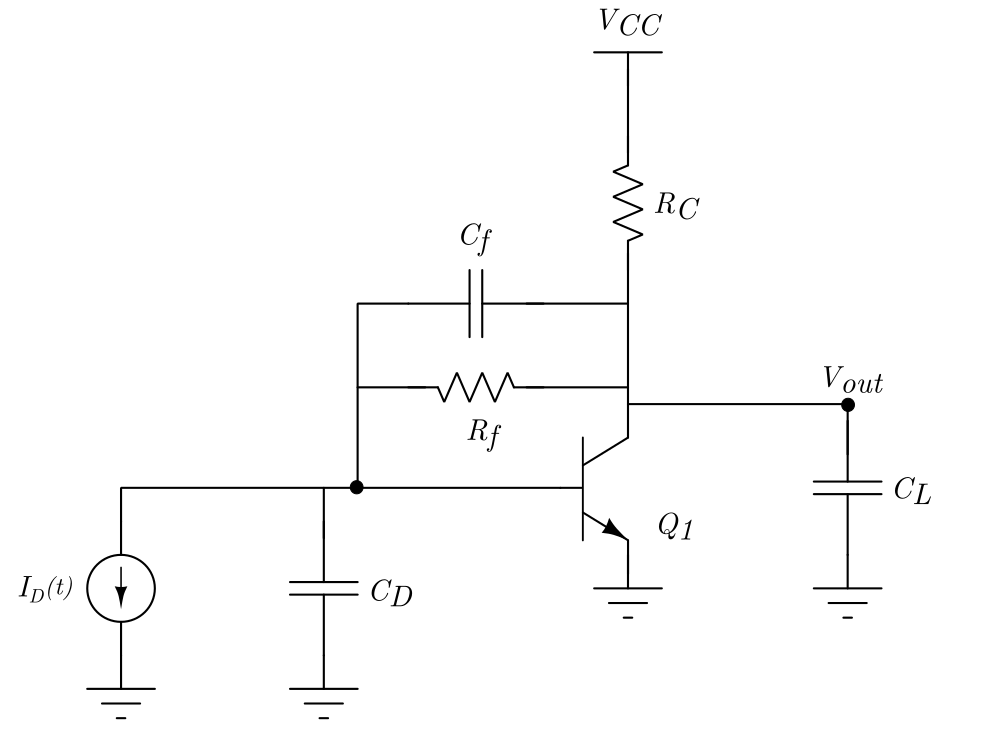}
	\includegraphics[scale=0.39]{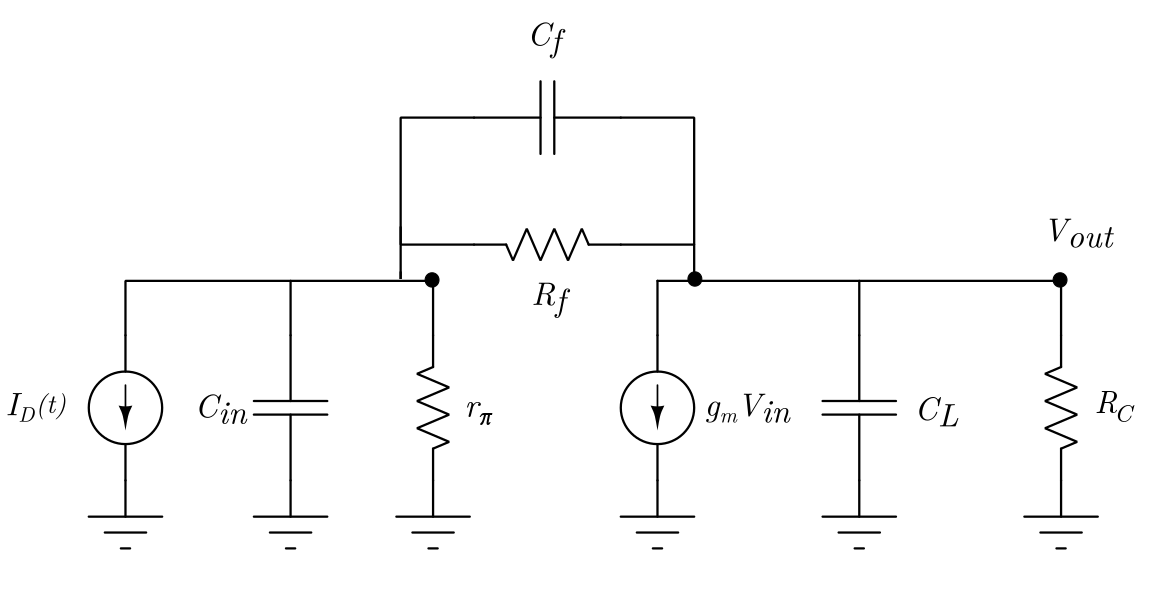}
	\caption{\footnotesize{Schematic of the FB-TIA self-biased circuit} (left) and corresponding small signal model (right).}
	\label{fig:SiGe}
\end{figure}
A simple implementation of this solution is given by the so-called \textit{self-biased circuit} (Fig.~\ref{fig:SiGe}, left) \cite{Razavi}. This topology can be analyzed in detail \cite{LaiCossu} using the corresponding small signal model (Fig.~\ref{fig:SiGe}, right) and finding the following second order transfer function,
\begin{equation}
R_m(s) = \frac{R_{m_0}}{(1+s\tau)^2}.
\label{eq:RmFast}
\end{equation} 

When a current like the one shown in Fig.~\ref{fig:img3} is processed by such electronics, the output signal $V_{out}$ has the following characteristics,

\begin{equation}
T_{peak}=\frac{e^{\frac{t_c}{\tau}}t_c}{e^{\frac{t_c}{\tau}}-1},
\end{equation} 

and,

\begin{equation}\label{vpeak_values}
V_{peak}=I_0R_{m_0}e^{-\frac{T_{peak}}{\tau}}(e^{\frac{t_c}{\tau}}-1).
\end{equation} 

In our case, $I_0 = \frac{Q_{MVP}}{t_c} \sim 10~\mu A$, while $R_{m_0}$ is the DC trans-impedance introduced in section \ref{sec:2}, which depends on the resistances of the circuit and on the trans-conductance $g_m$ of the transistor,

\begin{equation}\label{key}
	R_{m_0}=\frac{r_\pi g_mR_C R_f-r_\pi R_C}{(R_f+R_C+r_\pi(1+g_mR_C)}.
\end{equation}

Thanks to the wide bandwidth of the Si-Ge in use, and the high value of the trans-conductance $g_m$, we see the dominant contribution to the noise is given by the fluctuation of the bias current $I_b$ at the base of the transistor, which can be written \cite{LaiCossu},

\begin{equation}
\sigma_{v,b} \sim \sqrt{\frac{k_BT}{4r_\pi}}\frac{R_{m_0} }{\sqrt{\tau}},
\end{equation}
where $r_\pi=\frac{V_T}{I_b}$ is the dynamic input resistance of the bipolar transistor, $V_T=\frac{k_B T}{e}$ is the thermal voltage and $k_B$ is the Boltzmann constant.

The maximum slope of the signal, when the time constant $\tau$ is smaller than the duration of the current pulse $t_c$, is equal to
\begin{equation}
	V'_{out}=\frac{I_0R_{m_0}}{\tau e}.
\end{equation}

The electronic jitter $\sigma_{ej}$ (Eq.\ref{jitter}) can therefore be written,
\begin{equation}\label{jitterib}
\sigma_{\rm ej,b}=\sqrt{\frac{k_BT}{ 4 r_\pi}}\frac{e}{I_0} \sqrt{\tau}.
\end{equation}

The electronic jitter $\sigma_{ej}$ increases with the square root of the time constant $\tau$, while the maximum possible value for $V_{peak}$ is given by $V_{peak}=I_0R_{m_0}$ and is reached ideally for $\tau=0$. We can now choose a time constant for our system (sensor+electronics) that allows the best signal to noise ratio (SNR), which shows a maximum when,
\begin{equation}\label{best_tau}
SNR_{max} = \Bigg(\frac{V_{peak}}{\sigma_{v,b}}\Bigg)_{max}\longrightarrow \tau \sim 0.335~t_c \approx 70 ~\text{ps}.
\end{equation}

Since Eq.~\ref{eq:RmFast} is the transfer function of a second order system, Eq. \ref{unipolar_bw} gives the $-3dB$ frequency as,

\begin{equation}
f_{-3dB}\sim \frac{0.54}{2 \pi\tau} = \frac{0.54}{2 \pi~ 0.335 ~ t_c} \approx \frac{0.26}{t_c},
\end{equation}

and using an average value for the charge collection time of $t_c=200~\text{ps}$,
\begin{equation}
f_{-3dB}\approx 1.3 ~\text{GHz}.
\end{equation}
Using this simplified model we find that the bandwidth optimising the SNR is smaller than the bandwidth of the signal, as defined in Eq.~\ref{sigBW}. On the other hand, we are still neglecting the contribution of the other noise sources such as the resistances $R_f$, $R_C$ and the fluctuations of the bias current $I_C$. Taking these effects into account, the best time constant will have a slightly higher value than given by Eq.~\ref{best_tau}. We must also consider that the current duration is not constant, as shown by the CCT distribution (Fig.~\ref{fig:img2}). In the case of signals that are longer than the average time $t_c$, too short of a time constant $\tau$ leads to integration of a smaller proportion of the total charge and, according to Eq. \ref{vpeak_values}, leads to an output signal of lower amplitude. To understand the performance achievable by this configuration we need to consider realistic values for the trans-conductance $g_m$, the resistances of the circuit and the supply voltage $V_{CC}$ and find the best value for the trans-impedance $R_{m_0}$ that leads to the best time constant $\tau$.

\subsubsection{Theoretical performance estimate}\label{teoretical}
To fully exploit the speed of our 3D sensors we need a transistor with a high transition frequency $f_T$. HBT Si-Ge RF transistors can typically sustain bias currents $I_C$ of the order of a few tens of mA, which corresponds to values of trans-conductance $g_m$ close to unity. The values of the internal capacitance $C_{be}$ (base-emitter) and $C_{ce}$ (collector-emitter) are smaller than 1~pF, while the capacitance $C_{cb}$ (collector-base) working in feedback in our common emitter configuration is smaller than 100~fF. It can be shown \cite{LaiCossu} that the time constant of the chosen circuit can be written as,
\begin{equation}\label{tau_FT}
	\tau=\sqrt{\frac{R_fR_C r_\pi \xi}{r_\pi(1+g_mR_C)+R_C+R_f}} \approx \sqrt{\frac{R_{m_0}\xi}{g_m}},
\end{equation}

with the quantity $\xi$ given by,
\begin{equation}\label{xidef}
	\xi = (C_LC_{in}+C_LC_f+C_{in}C_f),
\end{equation}
and contains all the capacitance involved in the circuit,
\begin{equation}
C_{in}=C_D+C_{be}~~~~~~~~C_f=C_{cb}~~~~~~~~~~C_L=C_{ce}.
\end{equation}

We are still not considering any parasitic capacitance nor load capacitance of the following stage. Having a fixed value for the capacitance means the problem to define the time constant $\tau$ is now related to the value of the power $V_{CC}$ and the resistances $R_f$ and $R_C$. The correct choice has to be made, taking into account the stability of the circuit that can be established considering the circuit damping factor $\zeta$, which is a function of its resistances, capacitance and transistor trans-conductance (see~\cite{LaiCossu} for a wide explanation). An under-damped system can lead to an unwanted oscillating behavior, while an over-dumped system might not exploit the full potential of the active component. A numerical example, with realistic values for all components and considering the performance of the best-in-class ultra low noise HBT transistors \cite{infineon}, can be found in \cite{LaiCossu}. This shows that with a single stage and an input charge $Q_{MPV}\sim 2~\text{fC}$, the output voltage can be of the order of
\begin{equation}
	V_{peak}\sim 10 ~\text{mV},
\end{equation}

while the time constant is about,

\begin{equation}
	\tau \sim 180 ~\text{ps}.
\end{equation}

This time constant corresponds to a trans-impedance $R_m(s)$ bandwidth $f_{-3dB}\sim 480~\text{MHz}$. The $V_{peak}$ value found, to be compared with a typical noise of a few hundreds of $\mu V$, also shows that it is opportune to add a second amplification stage in order to increase the amplitude of the signal. The choice in this case is to use the same self-biased trans-impedance configuration with the same ultra low noise Si-Ge transistor, optimizing the operating point and the size of the components in such a way to reach the best performance in terms of SNR and electronic jitter. A second stage further reduces the bandwidth, which can be exploited to optimize the overall time constant to limit ballistic deficit. With the data-sheet values of the components and the working point established in \cite{LaiCossu}, we find that the estimated electronic jitter can still be of the order of
\begin{equation}\label{jitter_est}
	\sigma_{ej}\sim 7~\text{ps}.
\end{equation}

The value of the jitter $\sigma_{ej}$ in Eq.~\ref{jitter_est} depends on the capacitance of the pixel $C_D$, which in the example was chosen to be $C_D\approx1~\text{pF}$. Also the transistor input capacitance, which is enhanced by the Miller effect,  has a typical value of the same order. For values of $C_D<1~\text{pF}$, we expect to observe a higher SNR and a lower jitter performance, while for values of $C_D$ significantly higher than 1~pF, both the SNR and the $\sigma_{ej}$ are expected to be worse.

\subsection{Spice simulations}
\begin{figure}[h] 
	\centering
	\includegraphics[width=0.92\textwidth]{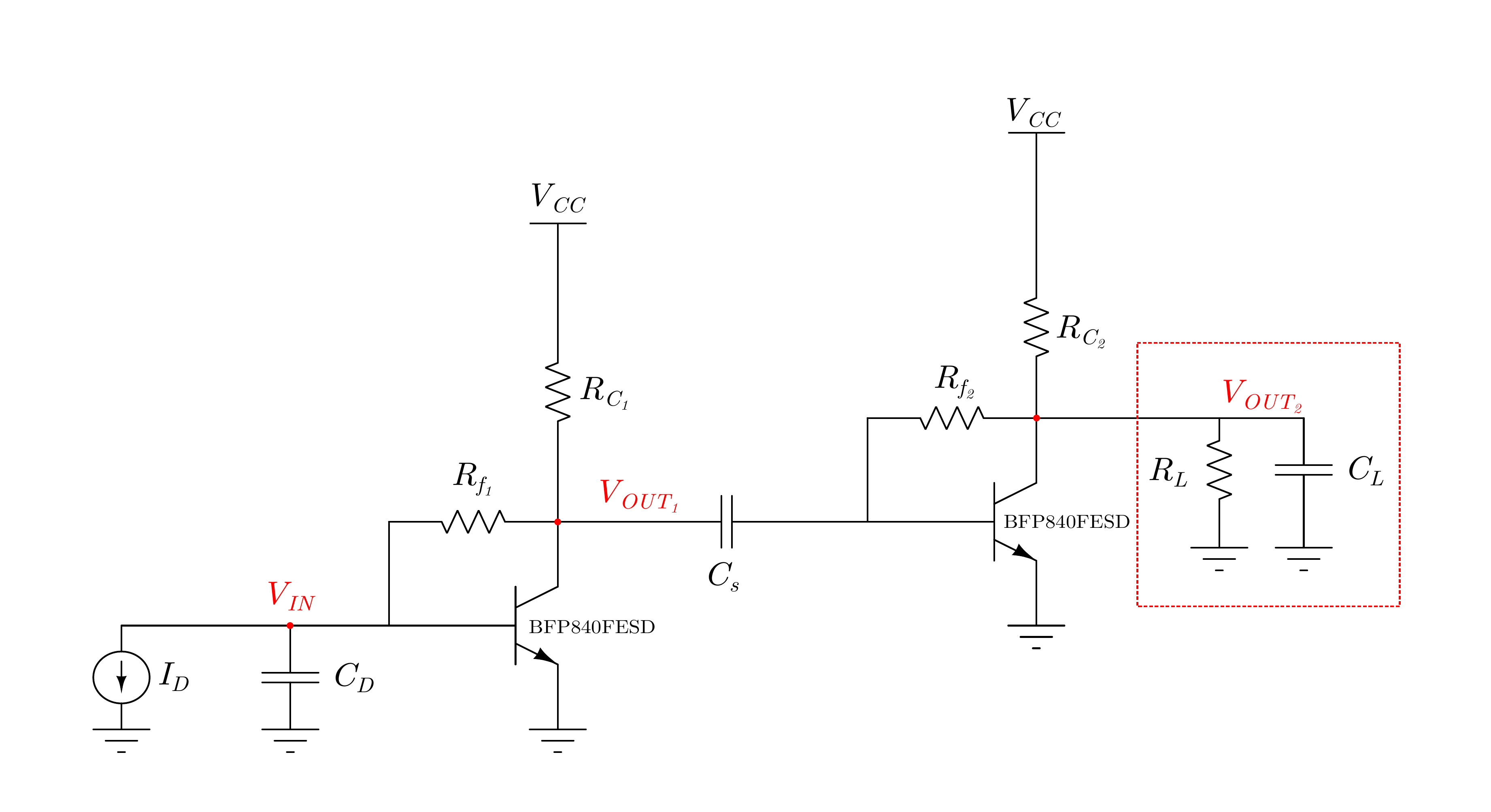}
 \vspace{-6mm}
	\caption{\footnotesize{Simplified schematic of the two stages circuit used for the electronics front-ends developed.}}
	\label{spice0}
\end{figure}

\begin{figure}[h] 
	\centering
 \includegraphics[scale=0.35]{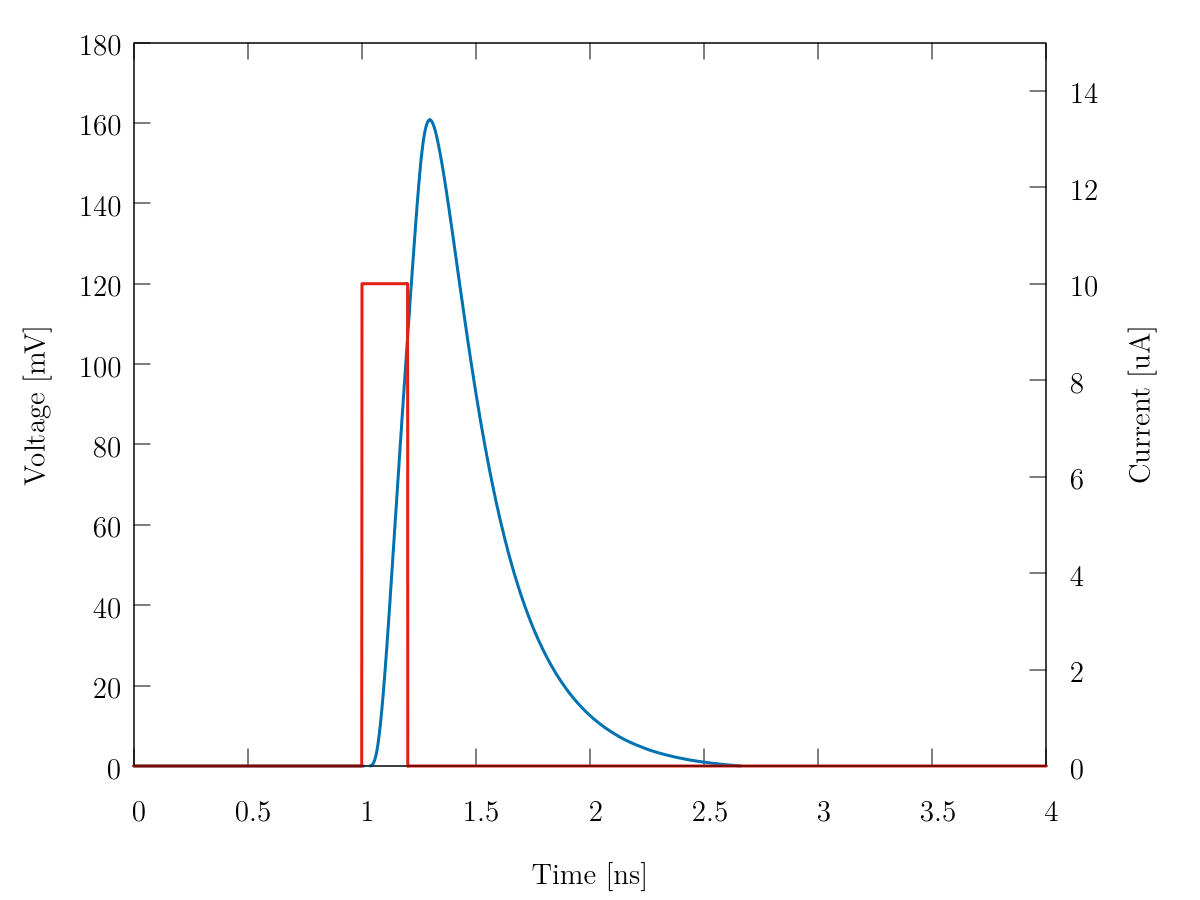}
  \vspace{-3mm}
	\caption{\footnotesize{Transient simulation of the two stages circuit for a rectangular input current as the one shown in Fig.~\ref{fig:img3} and corresponding output signal (blue waveform). The output signal has been reversed for better viewing.}}
	\label{spice1}
\end{figure}

\begin{figure}[h] 
	\centering
	\includegraphics[width=0.8\textwidth]{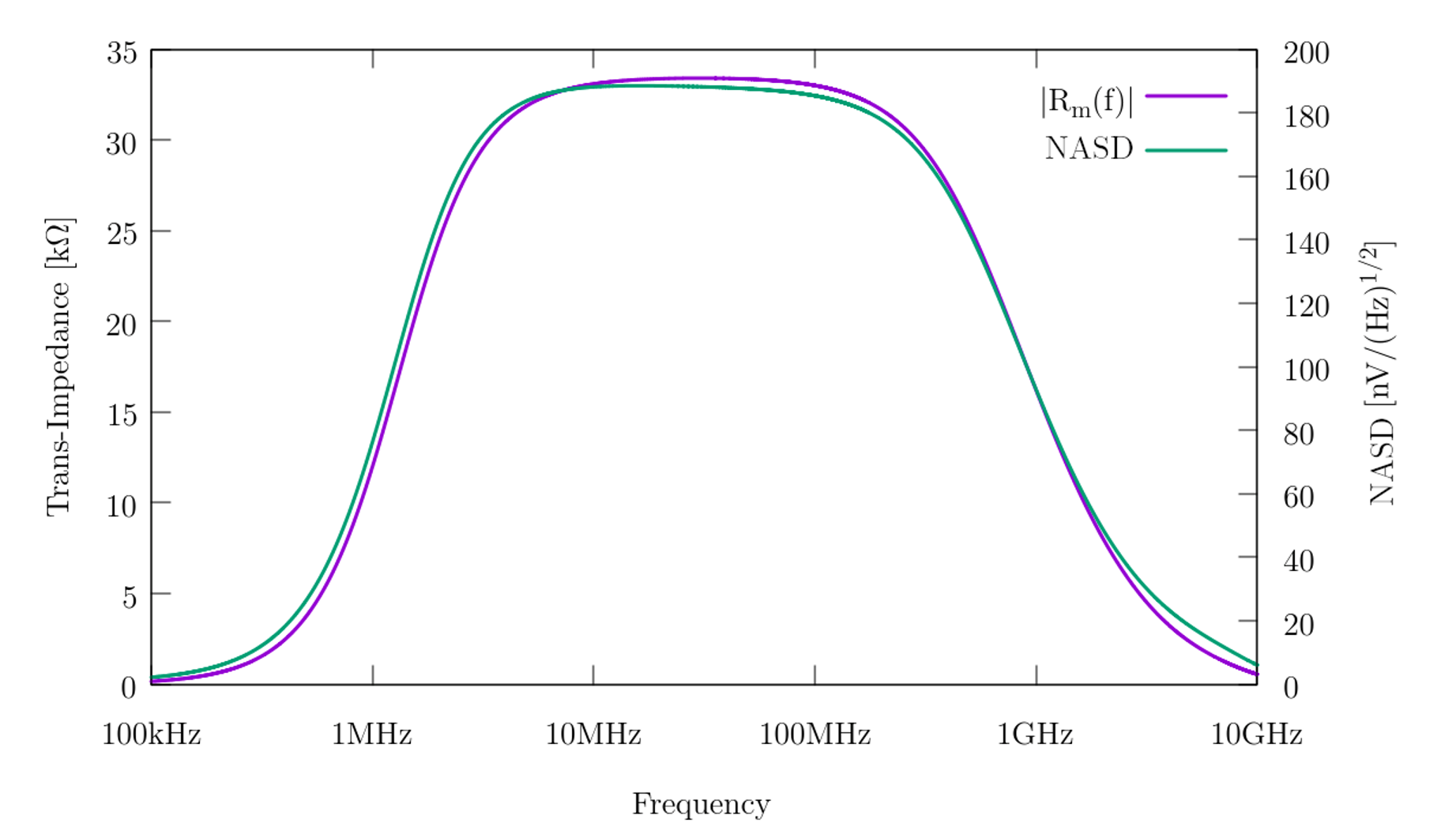}
  \vspace{-3mm}
	\caption{\footnotesize{Results of simulations in the frequency domain of the two stage board and noise analysis. The measured bandwidth is $f_{-3dB}\sim 780~\text{MHz}$ while the integrated voltage noise of the NASD at the output is $\sigma_v \sim 6.1~\text{mV}$. }}
	\label{spice2}
\end{figure}

Together with the analytic model shown in the previous section, Spice-based simulations are very useful to evaluate circuit performance. Component manufacturers usually provide Spice models that are suitable to perform transient, frequency, and noise analysis simulations. Fig.~\ref{spice1} shows an example of transient simulation of the two-stage circuit based on the self-biased topology (Fig.~\ref{spice0}).
These simulations can help to evaluate the operating point of the active components and the stability of the system. In particular, parametric simulations are very useful, where the effect of varying the value of a particular component or of the power supply can be investigated. Through AC simulations, it is possible to control the bandwidth of the front-end, while through noise analysis, it is possible to evaluate the Noise Amplitude Spectrum Density (NASD) produced by all noise sources (Fig.~\ref{spice2} green line ), such as Johnson-Nyquist noise, shot noise etc. The integral of the NASD gives the RMS value of the noise at the node of interest.
Considering the simulation illustrated in Fig.~\ref{spice1}, the output voltage shows $V_{peak} \sim 160~\text{mV}$, while with the noise analysis we find $\sigma_v \sim 6.1~\text{mV}$. The bandwidth is $f_{-3dB}\sim 780~\text{MHz}$ and the mid-band trans-impedance is $ R_{m_0}\sim 35~ k\Omega$. The obtainable electronic jitter $\sigma_{ej}$ is estimated using Eq. \ref{jitter}, where the slope at the chosen threshold is calculated on the signal obtained with the transient simulation, while the output noise, is taken from the result of the noise analysis. The estimated electronic jitter $\sigma_{ej}$ at the output was found to be $7~\text{ps}$. On the other hand, this simulation doesn't take into account any parasitics or losses.
The complexity of the simulation can be increased, modeling the effect of parasitic capacitance and transmission lines, so as to gain indications on the PCB design and signal integrity, which is of particular importance at high frequencies. An accurate evaluation of these effects is decisive for the success of the design implementation, and the results must be correctly interpreted so as to not overestimate the performance of the system. 

\section{Electronics boards}
\label{sec:5}
This section shows the timing front-end electronics fabricated for the TimeSPOT 3D sensors. Several single channel versions have been designed, from the first prototypes to later versions, optimized to reduce high frequency losses and improve stability. In addition, a four-channel version has been produced, which allows characterization of neighboring sensors of the same test structure for charge sharing studies.

\begin{figure}[h] 
	\centering
	\includegraphics[width=0.98\textwidth]{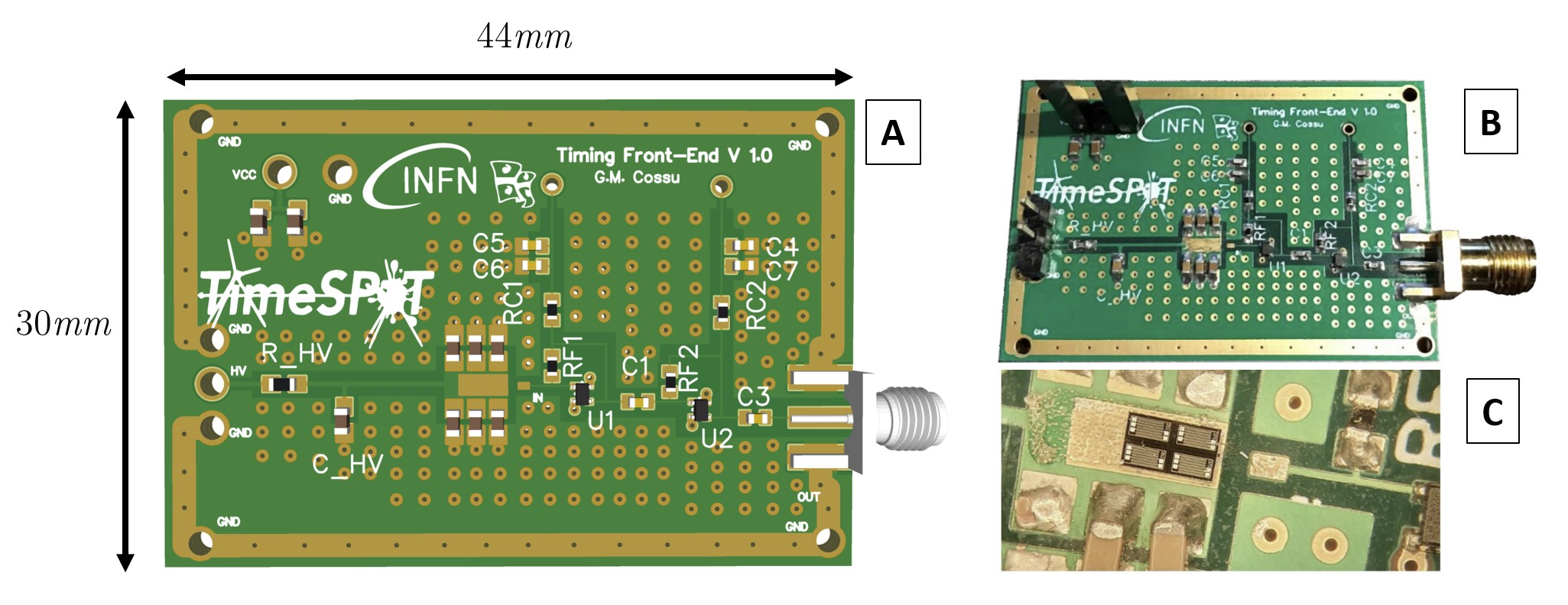}
	\caption{\footnotesize{A) Rendering of the of the single channel board \textbf{TFE v1.0}; B) One of the boards produced and assembled; C) Test structure with several 3D TimeSPOT sensors bonded to the electronics input pad. }}
	\label{board1}
\end{figure}

\subsection{Single channel boards}
The first single channel board, \textbf{TFE v1.0}, is shown in Fig.~\ref{board1}. It consists of a double TIA stage like the one described in the previous section, based on the Si-Ge ultra low noise transistors produced by Infineon \cite{infineon}. The board is equipped with a 4.0x4.0 mm$^2$ pad for sensor housing. The usable pad surface for gluing the sensor is 3.0x1.5 mm$^2$  due to the footprint of the bypass capacitors. The board allows the supply of the High Voltage bias of the sensor, which can be bonded to the input of the electronics channel by means of a 0.8x0.5 mm$^2$ pad. The PCB has a four layer stack-up and impedance-controlled tracks with $Z=50~\Omega$. The passive components of the front-end have packaging 0402 (1.0x0.5 mm$^2$) and the surface finish is ENIG (Electro-less Nickel Immersion Gold), which facilitates wire bonding.

\begin{figure}[h] 
	\centering
	\includegraphics[width=0.98\textwidth]{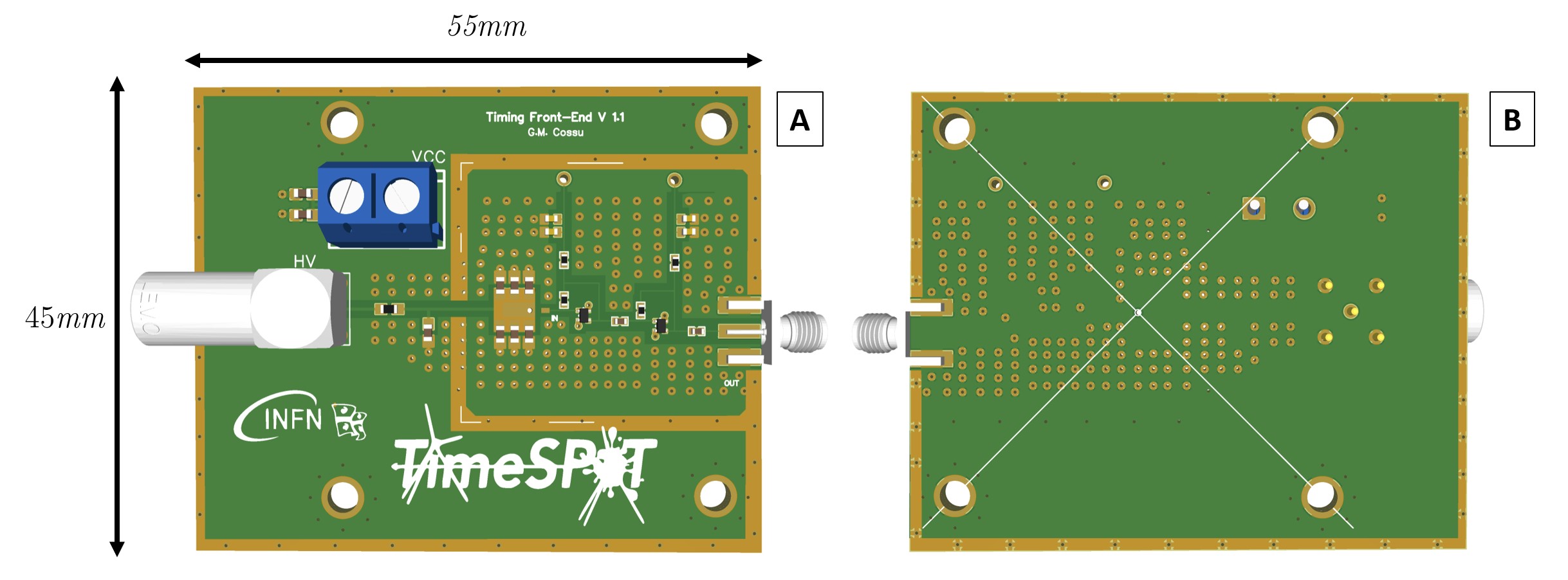}
  \vspace{-3mm}
	\caption{\footnotesize{A) Rendering of the of the single channel board \textbf{TFE v1.1}; B) Back of the PCB.}}
	\label{board2}
\end{figure}

Two later versions of the board were then developed. The second version \textbf{TFE v1.1} (Fig.~\ref{board2}), has the same layout for the front-end but a different size of the PCB which has been equipped with holes to facilitate mounting in measurement setups. The holes have been positioned such that the test structure (which normally has multiple sensors) is in the center of the symmetry axes (Fig.~\ref{board2} B). This allows rotation of the board by $90^{\circ}$ without the need for re-alignment, and gives more freedom in the creation of a measurement setup.

\begin{figure}[h] 
	\centering
	\includegraphics[width=0.98\textwidth]{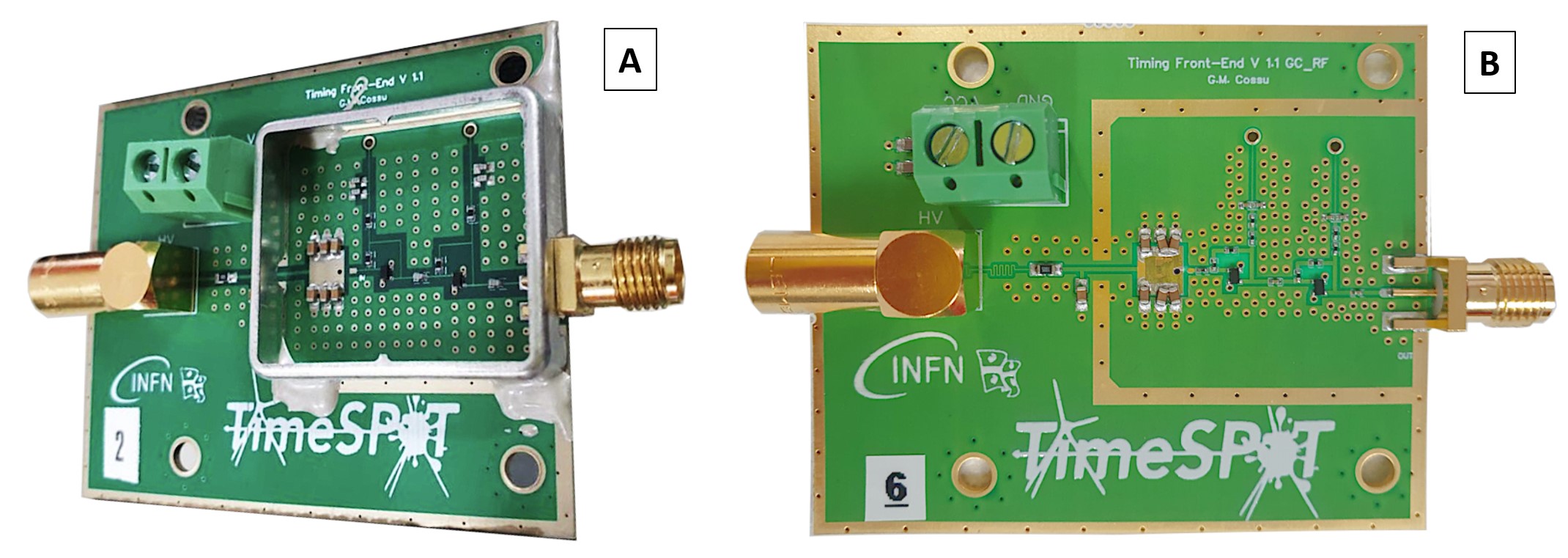}
 \vspace{-3mm}
	\caption{\footnotesize{A) The board \textbf{TFE v1.1} with the RF shield frame mounted; B) Version of the single channel \textbf{TFE v1.1 GC\_RF} made using ROGERS laminates and passive components with packaging 0201. }}
	\label{board3}
\end{figure}

 The sensor housing pad has a hole with a size of 500 $\mu$m, which can be enlarged from the bottom if necessary. This reduces the material with which a particle interacts. In addition, an RF shield has been added (Fig.~\ref{board3} A).
The third version \textbf{TFE v1.1 GC} has the same footprint as the \textbf{TFE v1.1} but the front-end has been carefully redesigned with the aim of preserving the high frequencies of the signal and limiting losses due to impedance mismatch (Fig.~\ref{board3} B). Passive components with  packaging  0201 (0.3x0.6 mm$^2$) were used and the stack-up was modified to eliminate track width changes. A fourth version of the board \textbf{TFE v1.1 GC\_RF} was also produced using high frequencies substrates between the conductive layers of the PCB (i.e ROGERS RO4450 for the core and ROGERS RO4350 for the pre-preg) in order to have tight control on dielectric constants and maintain low losses at high frequency.

\subsubsection{Boards performance}

\begin{figure}[h] 
	\centering
	\includegraphics[width=0.65\textwidth]{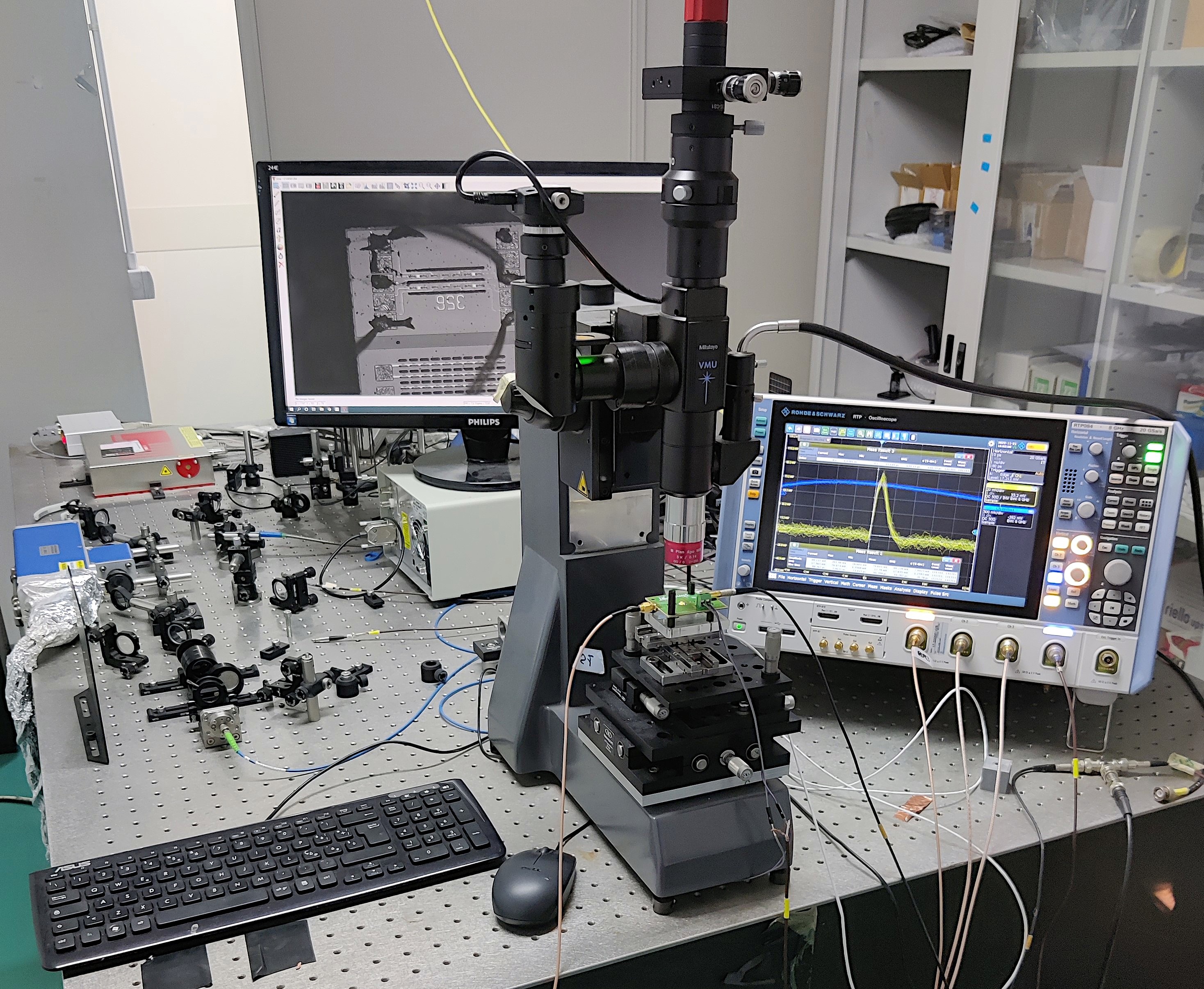}
	\caption{\footnotesize{ The laser based setup used to characterize the boards. }}
	\label{laser}
\end{figure}

\begin{figure}[h] 
	\centering
	\includegraphics[width=0.9\textwidth]{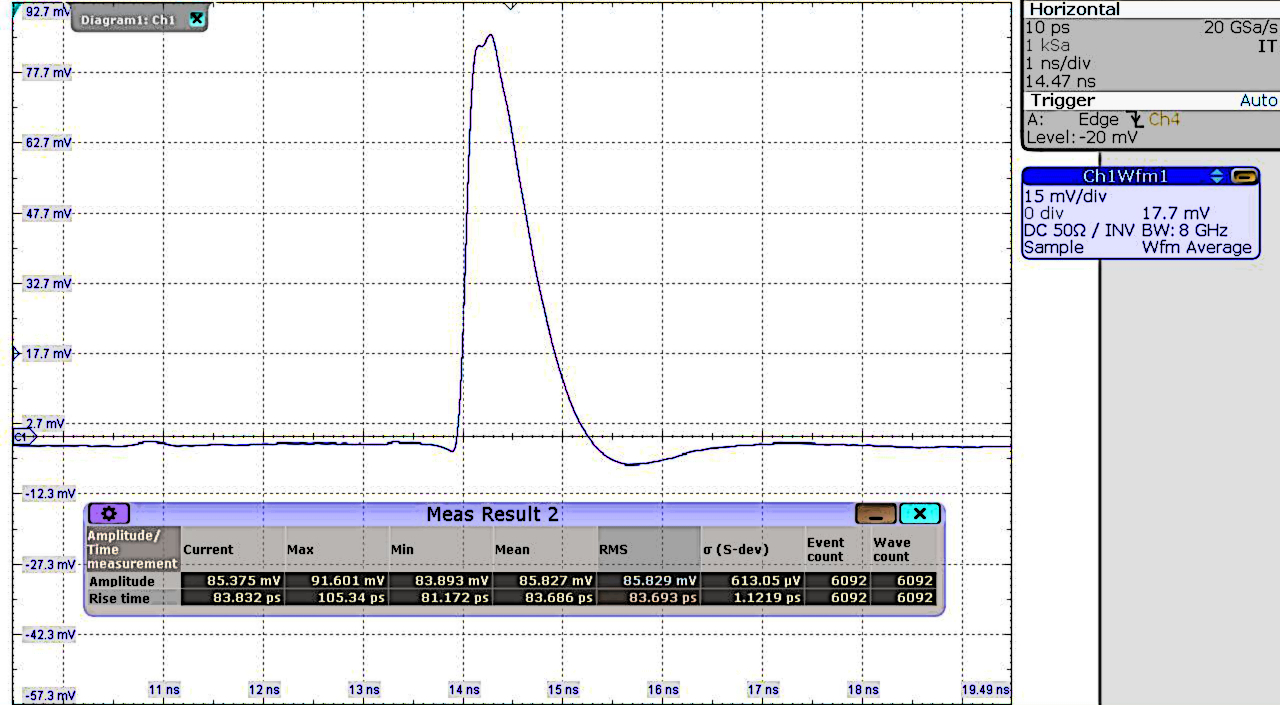}
 \vspace{-3mm}
	\caption{\footnotesize{ Average waveform of the board \textbf{TFE v1.1 GC\_RF}. }}
	\label{oscillo}
\end{figure}

\begin{figure}[h] 
	\centering
	\includegraphics[width=0.98\textwidth]{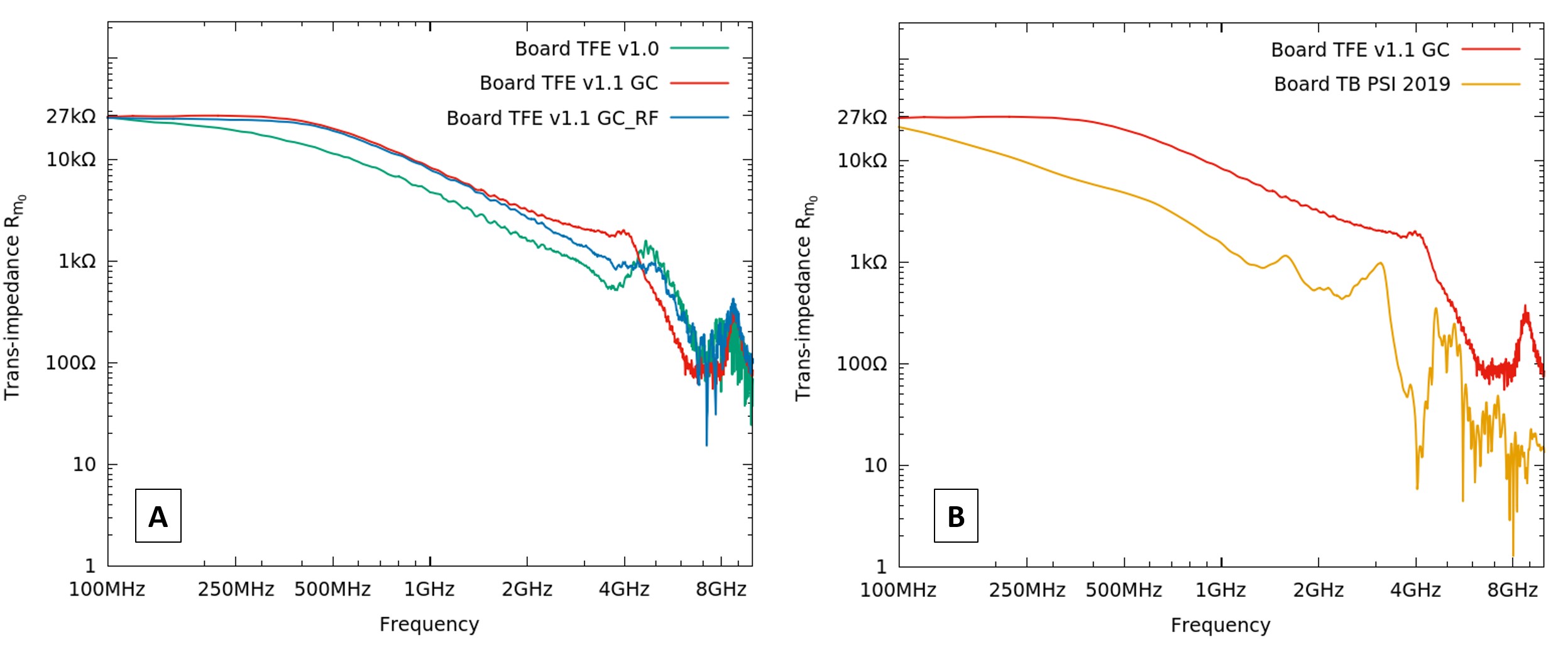}
 \vspace{-4mm}
	\caption{\footnotesize{A) Trans-impedance in the frequency domain of the three single channel boards produced with $C_D \sim 1\text{pF}$. B) Comparison of the trans-impedance in the frequency domain of (\textbf{TFE v1.1 GC}) with the board used in the test beam at PSI in \cite{JINST-TimeSpot}. }}
	\label{board6}
\end{figure}

Characterization of the boards was done using a setup equipped with an oscilloscope with 8 GHz analogue bandwidth and 20 GSa/s sampling, and a 200 fs pulsed, 1030 nm wavelength laser with a minimum spot size on its target of 5 $\mu m$ \cite{LabMeas} (Fig.\ref{laser}).
The amplitude, rise-time, noise and SNR values were measured directly using the measurement functions of the oscilloscope used. For time resolution measurements, a second board was used for time reference. This board houses another 3D sensor onto which a reflection of the laser is directed. Since the equivalent deposit on this time reference is about 10 MIPs, the electronic jitter is strongly reduced and a jitter of $907~\text{fs}$ is obtained. The electronic jitter $\sigma_{ej}$ of the DUT is therefore calculated as the standard deviation of the measurement of the delay between the signal from the DUT and the one of the time reference. 
As described in section \ref{sec:2}, the action of the system on the currents produced by the sensor can be represented by a transfer function $R_m(s)$. However, this function includes internal characteristics of both the electronics and the sensor, which means that different sensors connected to the same electronics correspond to different transfer functions $R_m(s)$. The type of connection (e.g. wirebonding), the presence of parasitic capacitance and the length of the transmission line can also influence the trans-impedance $R_m(s)$.
An accurate trans-impedance measurement therefore requires the sensor to be connected to the electronics. For this reason the trans-impedance in the frequency domain (Fig.~\ref{board6}) was calculated by following a semi-empirical method~\cite{3D-accurate}. This method consists in pulsing the sensor with the laser and measuring the average output signal (Fig. \ref{oscillo}). By means of a simulation with the TCode package ~\cite{Loi_2021}, the current corresponding to the deposition of the laser in the chosen position is obtained. Finally the trans-impedance of the system is obtained by means of a deconvolution made with the software TFBoost~\cite{Tfboost}.

The measured values for the main quantities of the different boards are summarized in Table \ref{table1}, which shows the values of integrated noise $\sigma_v$, the SNR, the rise time $t_r$ from 20\% to 80\% of the $V_{peak}$, and the jitter $\sigma_{ej}$ corresponding to the MPV of charge deposit of a MIP in the sensor ($Q_{MPV}=2fC$). The precision on the values shown in the tables is of the order of 1\%. The amplitude for each board corresponding to a MIP deposit was calibrated considering the results obtained from measurements made during several test-beams \cite{CardiniPisa} \cite{Andreaiworid} and represents a typical value with a voltage bias of 100V.

\begin{table}[h] 
\centering
\setlength{\tabcolsep}{10pt}
\renewcommand{\arraystretch}{1.20}
\begin{tabular}{|c|c|c|c|}
\hline
\textbf{Board} & \textbf{TFE v1.1} *  & \textbf{TFE v1.1 GC } * & \textbf{TFE v1.1 GC\_RF} ** \\
\hline
\text{ $\sigma_v~~[\text{mV}]$}  & $2.5$  & $3.5$  & $3.3$  \\  
\hline
 \text{$V_{peak}~[\text{mV}]$} &  $50$   & $70$ & $80$ \\  
\hline
\text{ SNR } & $20$  & $20$ & $23$ \\  
\hline
\text{ $t_r~[\text{ps}]$ } & $105$  & $95$ & $80$ \\  
\hline
\text{$\sigma_{ej}~[\text{ps}]$}  & $7$  & $6$ & $6$ \\
\hline
 \footnotesize{ Power Consumption } $[\text{mW}]$ & $70$  & $70$ & $70$ \\
\hline
\end{tabular}

	\caption{\footnotesize{Main values of the boards measured using the setup described in \cite{LabMeas}. \footnotesize{* Values with a strip of ten sigle pixel\\
** Values with a single pixel. The precision on the values shown is of the order of 1\%. } }}
	\label{table1}
    \renewcommand\thetable{1}
\end{table}  

The boards versions \textbf{TFE v1.1} and \textbf{TFE v1.1GC} showed major differences. In particular, the layout optimization and the use of passive components of smaller packaging allowed a considerable increase of the front-end bandwidth. A trans-impedance of $~27~k\Omega$ was obtained with a power bandwidth higher than 500 MHz when connected to a strip sensor of ten pixel (Fig.~\ref{board6}.A). The increase in bandwidth is also visible from the values reported in Table \ref{table2}. This allows a larger signal amplitude, while maintaining the same SNR.  

\begin{table}[h] 
\centering
\setlength{\tabcolsep}{10pt}
\renewcommand{\arraystretch}{1.20}
\begin{tabular}{|c|c|c|c|}
\hline
\textbf{Board} & \textbf{TFE v1.1}  & \textbf{TFE v1.1 GC }& \textbf{TFE v1.1 GC\_RF} \\
\hline
~~~~~Power Bandwidth~~~~~& $326~\text{MHz}$  & $526~\text{MHz}$   & $515~\text{MHz}$ \\  
\hline
{Bandwidth} & $745~\text{MHz}$ & $1.02~\text{GHz}$ & $980~\text{MHz}$ \\  
\hline
Mid-band $R_{m_0}$ & $26k\Omega$  & $27k\Omega$ & $26k\Omega$\\  
\hline
\end{tabular}
	\caption{\footnotesize{Frequency characteristic of the pulse response of the boards obtained with the semi-empirical method and with a strip detector of 10 pixel ($C_D \sim 1pF$)}}
	\label{table2}
    \renewcommand\thetable{1}
\end{table}  
The version of the board made with ROGERS laminate has shown excellent performance in terms of bandwidth, especially with lower capacity sensors such as the single pixel. It is also slightly more sensitive to external noise sources and requires more accurate shielding. Fig.~\ref{board6}~B shows a comparison of the board \textbf{TFE v1.1 GC} with the one used at the PSI test beam in 2019~\cite{JINST-TimeSpot}, where the measured electronic jitter was $\sigma_{ej}\sim 15~\text{ps}$. The improvement with the \textbf{TFE v1.1 GC} board in terms of bandwidth and response uniformity over the whole spectrum is remarkable. In addition to this, it exhibits greater stability which has been tested with several sensors with different capacitance values, and always provides excellent performance.

\subsection{Multi-channel board}
\begin{figure}[h] 
	\centering
	\includegraphics[width=0.95\textwidth]{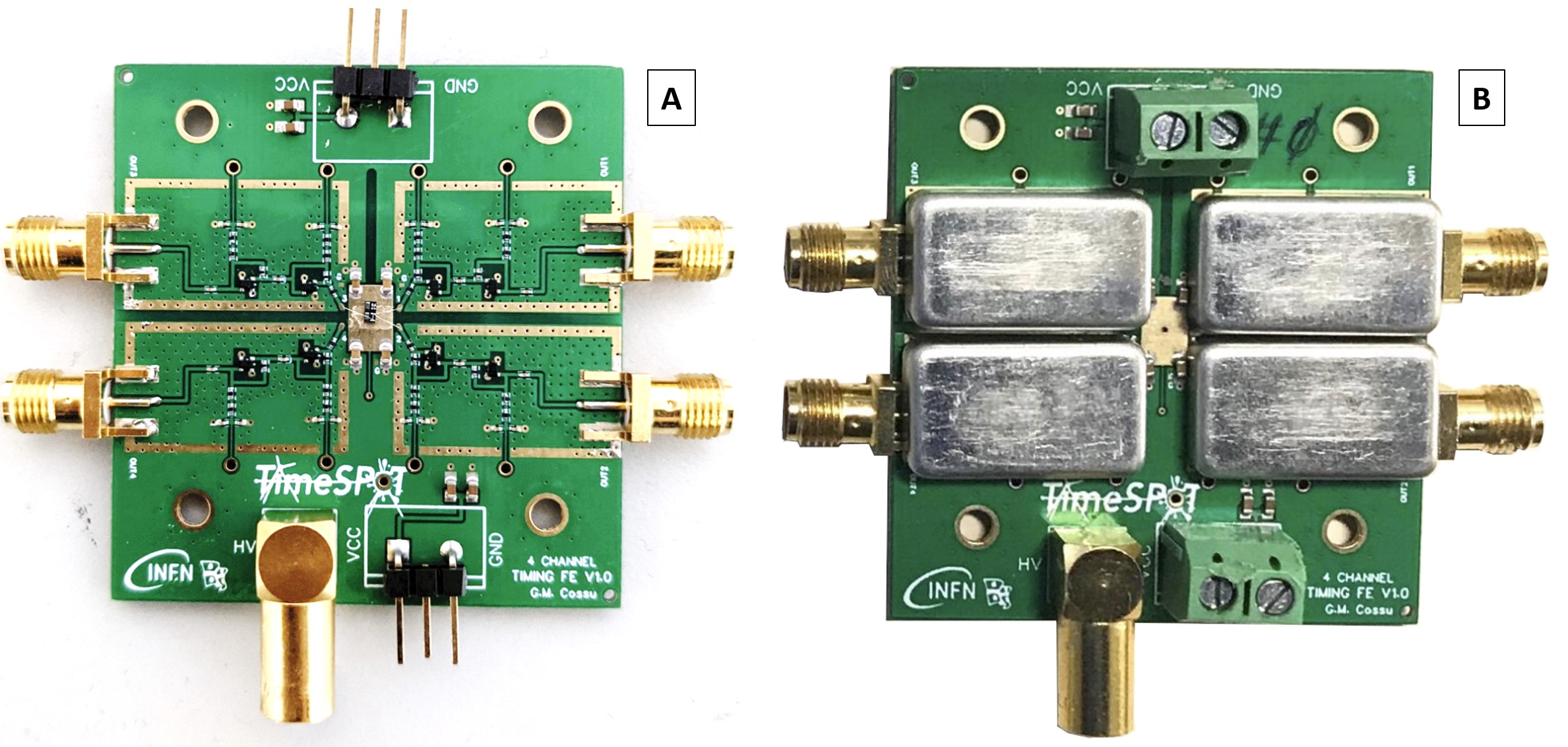}
	\caption{\footnotesize{A) The \textbf{TFE4CH v1.0} board with the test structure glued to the center of the pad. The wire-bonding to the individual electronics channels are visible. B) The RF shields designed for the individual front-ends. }}
	\label{board5}
\end{figure}

The four-channel version of the board \textbf{TFE4CH v1.0} is shown in Fig.~\ref{board5}. It was designed to characterize the timing performance of TimeSPOT sensors, taking into account the charge sharing between adjacent pixels.
The test structure that is positioned in the central pad must now be connected to different electronics channels that need to be properly decoupled to minimize cross-talk. Therefore, particular care has been taken in designing the ground layer in order to allow the current from the sensor to have the correct return path towards the relevant electronics channel.

The power layer has also been optimized, with two separate power supplies used for the upper and lower channels of the board. The optional use of individual RF shields for the electronics channels has also been provided (Fig.~\ref{board5}~B).
The size of the PCB is similar to that of the single channel board and the same principle of symmetrical holes has been used with respect to the sensor pad which makes it particularly easy to use in a measurement set-up. 
\begin{figure}[h] 
	\centering
	\includegraphics[width=0.99\textwidth]{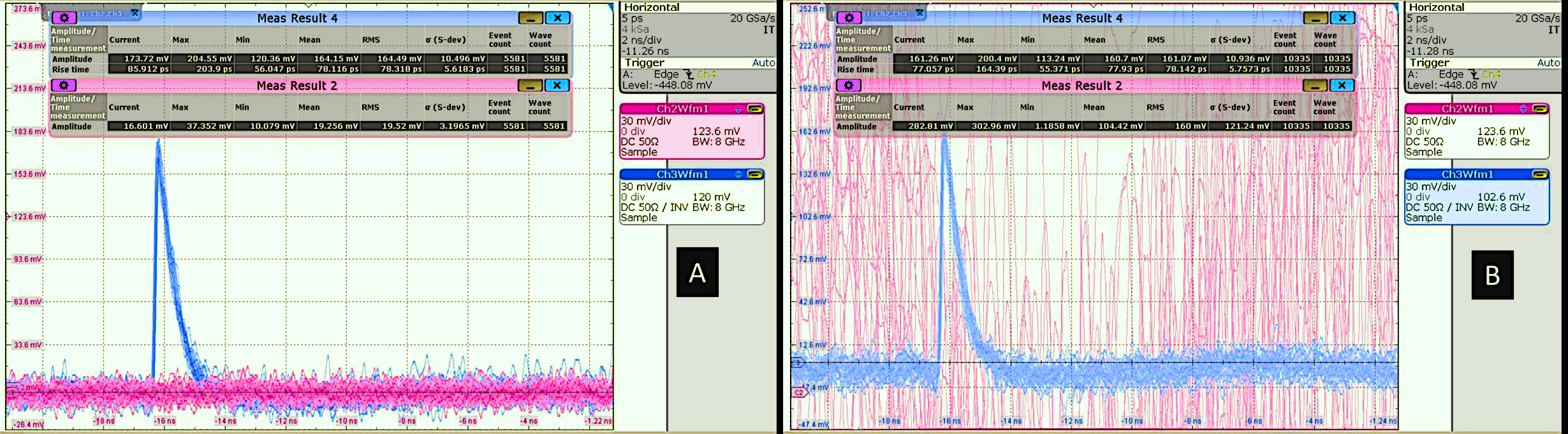}
	\caption{\footnotesize{ Signals acquired with the laser setup using the multichannel board with a test structure with two single pixels: Fig. A) The blue signal represents the pixel stimulated with the laser while the magenta signal represents that of the adjacent channel. Fig B) If a high amplitude noise is produced in the adjacent channel, no significant effect is noticed in the channel pulsed with the laser. }}
	\label{cross}
\end{figure}
The board showed excellent decoupling between channels with negligible cross-talk. This can be seen from Fig. \ref{cross} A) and B), which show the signals acquired on two adjacent channels of the multi channel connected to two single pixels of the same test structure. By injecting a high amplitude noise into one of the channels (simply touching the input of the transistor with a metal object), we have no effect in the channel pulsed with the laser. The bandwidth of the multichannel has not yet been calculated with the semi-empirical method described in section \ref{sec:5}, but oscilloscope measurements (Fig. \ref{cross}) show that the risetime and signal shape agree with what is obtained with the single channel version.

\section{Applications}
\label{sec:6}
The boards described were designed and assembled between 2019 and 2020. In the past years they have been used extensively to carry out many types of measurements. This section summarizes some of the applications of the produced boards. The first application was the use of the board in a setup for a complete in-laboratory characterization of 3D trench-type silicon pixel sensors. The setup is described in detail in~\cite{LabMeas}\cite{Laser}\cite{AndreaTrento} and one of the key features is to provide a time reference for time resolution measurements with the impressive resolution of $\sigma_t=907~\text{fs}$. This result was possible thanks to higher charge deposit (about 10 MIPs) and the property of 3D sensors to behave as ideal sensors  with  $\sigma_{sens} \approx 0$ when stimulated in the same position, leaving only $\sigma_{ej}$ as contribution to the time resolution. 

\begin{figure}[h] 
	\centering
	\includegraphics[width=0.65\textwidth]{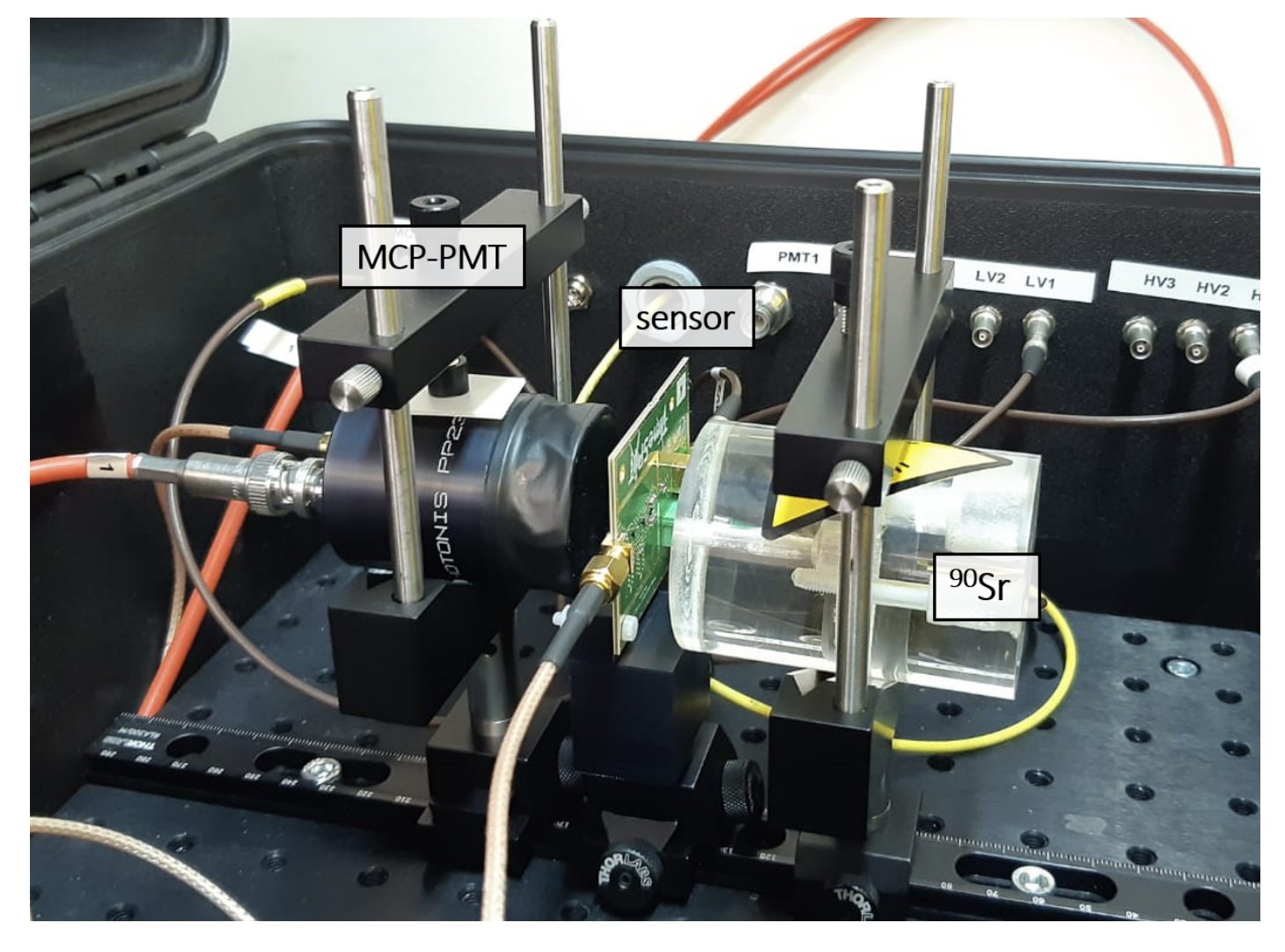}
	\caption{\footnotesize{The radioactive source setup based on the board developed (Taken from \cite{MichelaTrento}).}}
	\label{source_setup}
\end{figure}
Another application was to measure the time resolution of the 3D TimeSPOT sensor using a  $^{90}Sr$ radioactive source set-up~\cite{MichelaTrento} (Fig.~\ref{source_setup}). Recent results \cite{CardiniPisa} \cite{Andreaiworid}, obtained with an intensive measurement campaign with non-irradiated and irradiated 3D sensors at the SPS test-beam facility, have shown a temporal resolution of the detector + electronics system of about 11ps, which is about half of the value obtained in 2019 in the first test-beam at PSI \cite{JINST-TimeSpot}. The highly optimized electronics contributed significantly to this result, making the 3D TimeSPOT sensor the best temporally performing solid-state detector (in terms of timing resolution) without a gain mechanism currently. The TFE boards are still in  heavy use in the characterization campaigns of TimeSPOT sensors and further results will be surely published in the near future.

\section{Conclusions}
In order to characterize the timing performance of 3D-trench sensors produced within the TimeSPOT collaboration, various Si-Ge transistor-based boards have been designed, assembled and tested. The boards are single channel and multi-channel two-stage TIA, optimized in order to provide the best performance, both with single 3D pixels and with several pixels (strips) read out together, corresponding to capacitance values up to some pF. 
The measured jitter of the read-out channel for a charge deposit of 2 fC is lower than 7~ps.

\acknowledgments

This work was supported by the Fifth Scientific
Commission (CSN5) of the Italian National
Institute for Nuclear Physics (INFN), within the Project TimeSPOT. 

The authors wish to warmly thank the friend and colleague Gianni Corradi, from INFN Laboratori di Frascati, for his useful technical hints and precious discussions about the PCB layout.

\bibliographystyle{ieeetr}  
\bibliography{bib} 

\begin{thebibliography}{10}

\bibitem{FTDR}
C.~M. LHCb~Collaboration, ``{Framework TDR for the LHCb Upgrade II $\text{-
  Opportunities in flavour physics, and beyond, in the HL-LHC era} $},'' tech.
  rep., CERN, Geneva, 2021.

\bibitem{Ferrero_2020}
M.~Ferrero {\em et~al.}, ``Evolution of the design of ultra fast silicon
  detector to cope with high irradiation fluences and fine segmentation,'' {\em
  Journal of Instrumentation}, vol.~15, pp.~C04027--C04027, apr 2020.

\bibitem{Iacobucci_2022}
G.~Iacobucci {\em et~al.}, ``Efficiency and time resolution of monolithic
  silicon pixel detectors in {SiGe} {BiCMOS} technology,'' {\em Journal of
  Instrumentation}, vol.~17, p.~P02019, feb 2022.

\bibitem{adriano_last}
A.~Lai, ``4d-tracking in the 10-ps range: A technological perspective,'' {\em
  Frontiers in Physics}, vol.~10, 2022.

\bibitem{3D}
S.~Parker {\em et~al.}, ``{3D- A proposed new architecture for solid-state
  radiation detectors},'' {\em NIM A395, 328-34}, 1997.

\bibitem{Parker}
{Parker, S.I. and Kok A., and Kenney C. and Jarron P. and Hasi J. and Despeisse
  M. and Da Via C. and Anelli G.}, ``{Increased speed: 3D silicon sensors; fast
  current amplifiers},'' {\em IEEE Trans. Nucl. Sci., 58 (2), pp. 404-417},
  2011.

\bibitem{timespot}
{TimeSPOT collaboration}, ``web site: https://web.infn.it/timespot/index.php,''
  2020.

\bibitem{Piccolo_2022}
L.~Piccolo, S.~Cadeddu, L.~Frontini, A.~Lai, V.~Liberali, A.~Rivetti, and
  A.~Stabile, ``First measurements on the timespot1 {ASIC}: a fast-timing,
  high-rate pixel-matrix front-end,'' {\em Journal of Instrumentation},
  vol.~17, p.~C03022, mar 2022.

\bibitem{blum2008particle}
W.~Blum, W.~Riegler, and L.~Rolandi, {\em Particle Detection with Drift
  Chambers}.
\newblock Particle Acceleration and Detection, Springer Berlin Heidelberg,
  2008.

\bibitem{LAI2020164491}
A.~Lai {\em et~al.}, ``{First results of the TIMESPOT project on developments
  on fast sensors for future vertex detectors},'' {\em Nuclear Instruments and
  Methods in Physics Research Section A: Accelerators, Spectrometers, Detectors
  and Associated Equipment}, vol.~981, p.~164491, Nov. 2020.

\bibitem{JINST-TimeSpot}
L.~Anderlini, M.~Aresti, A.~Bizzeti, M.~Boscardin, A.~Cardini, G.-F.
  Dalla~Betta, M.~Ferrero, G.~Forcolin, M.~Garau, A.~Lai, A.~Lampis, A.~Loi,
  C.~Lucarelli, R.~Mendicino, R.~Mulargia, M.~Obertino, E.~Robutti, S.~Ronchin,
  M.~Ruspa, and S.~Vecchi, ``{Intrinsic time resolution of 3D-trench silicon
  pixels for charged particle detection},'' {\em {Journal of Instrumentation}},
  vol.~15, p.~P09029, {2020}.

\bibitem{3D-accurate}
D.~Brundu, A.~Cardini, A.~Contu, G.~Cossu, G.-F. Dalla~Betta, M.~Garau, A.~Lai,
  A.~Lampis, A.~Loi, M.~Obertino, G.~Siddi, and S.~Vecchi, ``{Accurate
  modelling of 3D-trench silicon sensor with enhanced timing performance and
  comparison with test beam measurements},'' vol.~16, p.~P09028, Sept. 2021.

\bibitem{3Dstepper}
M.~Boscardin {\em et~al.}, ``Advances in 3d sensor technology by using stepper
  lithography,'' {\em Frontiers in Physics}, vol.~8, 01 2021.

\bibitem{forcolin}
G.~Forcolin, ``Development of 3d trenched-electrode pixel sensors with improved
  timing performance.'' Hiroshima Symposium 2019, 2019.

\bibitem{Loi_2021}
A.~Loi, A.~Contu, and A.~Lai, ``Timing optimisation and analysis in the design
  of 3d silicon sensors: the {TCoDe} simulator,'' {\em Journal of
  Instrumentation}, vol.~16, pp.~P02011--P02011, feb 2021.

\bibitem{infineon}
{{Infineon Technologies}}, ``{Ultra low-noise Si-Ge:C transistors for use up to
  12 Ghz},'' 2022.

\bibitem{Benoit2016100}
M.~Benoit, R.~Cardarelli, S.~D{\' e}bieux, Y.~Favre, G.~Iacobucci, M.~Nessi,
  L.~Paolozzi, and K.~Shu, ``100 ps time resolution with thin silicon pixel
  detectors and a {SiGe} {HBT} amplifier - {IOPscience},'' {\em Journal of
  Instrumentation}, vol.~11, mar 9 2016.

\bibitem{Berretti_2017}
M.~Berretti, R.~Arcidiacono, E.~Bossini, M.~Bozzo, N.~Cartiglia, M.~Ferrero,
  V.~Georgiev, T.~Isidori, R.~Linhart, N.~Minafra, M.~Obertino, V.~Sola, and
  N.~Turini, ``Test of ultra fast silicon detectors for the {TOTEM} upgrade
  project,'' {\em Journal of Instrumentation}, vol.~12, pp.~P03024--P03024, mar
  2017.

\bibitem{Minafra_2020}
E.~Bossini and N.~Minafra, ``Diamond detectors for timing measurements in high
  energy physics,'' {\em Frontiers in Physics}, vol.~8, 2020.

\bibitem{UCSC}
{UCSC board Twiki page},
  ``https://twiki.cern.ch/twiki/bin/view/main/ucscsinglechannel,'' 2020.

\bibitem{CARTIGLIA201783}
N.~Cartiglia {\em et~al.}, ``Beam test results of a 16ps timing system based on
  ultra-fast silicon detectors,'' {\em Nuclear Instruments and Methods in
  Physics Research Section A: Accelerators, Spectrometers, Detectors and
  Associated Equipment}, vol.~850, pp.~83--88, 2017.

\bibitem{Razavi}
B.~Razavi, {\em {Design of Integrated Circuits for Optical communications}}.
\newblock John Wiley \& Sons, 2012.

\bibitem{LaiCossu}
A.~Lai and G.~M. Cossu, ``{High-resolution timing electronics for fast pixel
  sensors},'' {\em {arXiv2008.09867}}, 2020.

\bibitem{LabMeas}
M.~Aresti, A.~Cardini, A.~Lai, A.~Loi, G.~M. Cossu, M.~Garau, A.~Lampis,
  G.-F.~D. Betta, and G.~Forcolin, ``{Laboratory Characterization of Innovative
  3D trench-design Silicon Pixel Sensors Using a Sub-Picosecond Precision
  Laser-Based Testing Equipment},'' in {\em 2020 IEEE Nuclear Science Symposium
  and Medical Imaging Conference (NSS/MIC)}, pp.~1--6, 2020.

\bibitem{Tfboost}
D.~Brundu, A.~Contu, G.~M. Cossu, and A.~Loi, ``{Modeling of Solid State
  Detectors Using Advanced Multi-Threading: The TCoDe and TFBoost Simulation
  Packages},'' {\em Frontiers in Physics}, vol.~10, 2022.

\bibitem{CardiniPisa}
A.~Cardini, ``{10 ps timing with 3D trench silicon pixel sensors},'' 2022.
\newblock Talk given at the 15th Pisa Meeting on Advanced Detector.

\bibitem{Andreaiworid}
A.~Lampis, ``{10 ps timing with 3D trench silicon pixel sensors},'' 2022.
\newblock Talk given at the 23rd International Workshop on Radiation Imaging
  Detectors.

\bibitem{Laser}
M.~Aresti, A.~Cardini, G.~M. Cossu, M.~Garau, A.~Lai, A.~Lampis, and A.~Loi,
  ``{A Sub-Picosecond Precision Laser-Based Test Station for The Measurement of
  Silicon Detector Timing Performances},'' in {\em 2020 IEEE Nuclear Science
  Symposium and Medical Imaging Conference (NSS/MIC)}, pp.~1--4, 2020.

\bibitem{AndreaTrento}
A.~Lampis, ``{Sub-pixel characterization of innovative 3D trench-design silicon
  pixel sensors using ultra-fast laser-based testing equipment},'' 2021.
\newblock Talk given at the 16th Trento Workshop On Advanced Silicon Radiation
  Detectors.

\bibitem{MichelaTrento}
M.~Garau, ``{Laboratory characterization of 3D-trench silicon pixel sensors
  with a 90Sr radioactive source},'' 2021.
\newblock Talk given at the 16th Trento Workshop On Advanced Silicon Radiation
  Detectors.

\end{thebibliography}

\end{document}